\documentclass[sigconf]{acmart}
\usepackage{enumitem}
\newlist{questions}{enumerate}{2}
\setlist[questions,1]{label=\textbf{RQ\arabic*.},ref=\textbf{RQ\arabic*}}
\setlist[questions,2]{label=(\alph*),ref=\thequestionsi(\alph*)}

\usepackage{booktabs}
\usepackage{graphicx}
\usepackage{multirow}
\usepackage{tcolorbox}
\usepackage{xspace}

\newcommand{\TaSoBdAck}{This work is partially supported by Territori Aperti a project funded by Fondo Territori Lavoro e Conoscenza CGIL CISL UIL and by SoBigData-PlusPlus H2020-INFRAIA-2019-1 EU project, contract number 871042.}

\newcommand{\our}{MANILA\xspace}
\newcommand{\ourext}{Model bAsed developmeNt of ml pIpeLines with quAlity \xspace}

\AtBeginDocument{%
  \providecommand\BibTeX{{%
    \normalfont B\kern-0.5em{\scshape i\kern-0.25em b}\kern-0.8em\TeX}}}

\begin{document}

\title{Modeling Quality and Machine Learning Pipelines through Extended Feature Models}

\author{Giordano d'Aloisio}
\email{giordano.daloisio@graduate.univaq.it}
\orcid{0000-0001-7388-890X}
\affiliation{%
  \institution{University of L'Aquila}
  \city{L'Aquila}
  \country{Italy}
}
\author{Antinisca Di Marco}
\email{antinisca.dimarco@univaq.it}
\orcid{0000-0001-7214-9945}
\affiliation{%
  \institution{University of L'Aquila}
  \city{L'Aquila}
  \country{Italy}
}
\author{Giovanni Stilo}
\email{giovanni.stilo@univaq.it}
\orcid{0000-0002-2092-0213}
\affiliation{%
  \institution{University of L'Aquila}
  \city{L'Aquila}
  \country{Italy}
}

\renewcommand{\shortauthors}{d'Aloisio, et al.}

\begin{abstract}
  The recently increased complexity of 
  Machine Learning (ML) methods, led to the necessity to lighten both the research and industry development processes. 
  ML pipelines have become an essential tool for experts of many domains, data scientists and researchers, allowing them to easily put together several ML models to cover the full analytic process starting from raw datasets. Over the years, several solutions have been proposed to automate the building of ML pipelines, most of them focused on semantic aspects and characteristics of the input dataset. However, an approach taking into account the new quality concerns needed by ML systems (like fairness, interpretability, privacy, etc.) is still missing.  
  
  In this paper, we first identify, from the literature, key quality attributes of ML systems.
  Further, we propose a new engineering approach for quality ML pipeline by properly extending the Feature Models meta-model. The presented  approach allows to model ML pipelines, their quality requirements (on the whole pipeline and on single phases), and quality characteristics of algorithms used to implement each pipeline phase.  Finally, we demonstrate the expressiveness of our model considering the classification problem.
  
\end{abstract}



\maketitle

\section{Introduction}\label{sec:intro}

Machine Learning (ML) systems are increasingly becoming a used instrument, applied to all application domains and affecting our real life. Such systems can be defined as a set of one or more \emph{ML pipelines} that take as input raw (unprocessed) data and return actionable answers to questions in the form of machine learning models \cite{amershi2019software}.

Such pipelines usually require a good knowledge of the underlying ML domain to choose the best techniques and models to solve the targeted problem. For this reason, many methods have been developed in the last years to automate some phases of pipeline development \cite{RONKKO201513,6086455,HE2021106622}. However, these techniques are mainly focused on semantic characteristics of the input dataset, ignoring the new essential \emph{quality properties} introduced by such ML systems, such as dataset's bias reduction, model's interpretability, and fairness improvement \cite{muccini_software_2021,martinez-fernandez_software_2022,bosch_engineering_2021}.
Indeed, if we consider the impact that ML applications have in our lives, it is clear how assuring that these quality properties are satisfied is of paramount importance.
The importance of having \emph{high-quality} ML systems is also highlighted by some of the 17 sustainable development goals proposed by the United Nations \cite{united_nations_17_nodate}. In particular, to accomplish goals 5 (gender equality) or 10 (reduced inequalities) on a large scale, we will possibly rely on information systems. If those information systems include some ML pipelines (for classification or recommendation problems), it will be essential to properly manage and improve properties as \textit{Fairness} or \textit{Interpretability}.

In this paper, we present the first step on implementing \our (\ourext), a novel approach for engineering high-quality ML pipelines. First, we identify key quality attributes in ML systems by selecting the more adopted properties in the literature. Next, we discuss how ML pipelines can be modeled as Product Line Architectures \cite{capilla_overview_2014}, in which the variation points are the developed algorithms. For our aims, we use and extend the Feature Model formalism and meta-model\cite{kang1990feature}. The choice of which algorithms will be executed in the pipeline, case by case, is driven by the functional and quality requirements specified by the ML designer. 

In this work, we focus on the modeling and specification of ML pipelines with involved quality attributes. The pipelines are modeled through \textit{Quality and Feature Models}, which extend Feature Models \cite{kang1990feature} with quality properties of features. Models can also contain functional and quality requirements, specified by ML designers, that the approach will use to generate the final ML system. 

The \textit{Quality and Feature Models} are complaint to the \our \textit{Quality and Feature Meta-Model} (detailed in Section \ref{sec:meta}), that is an extension of the Feature Meta-Model \cite{burdek_reasoning_2016}.
In addition, on top of the defined meta-model, we implement a graphical editor which permits: \textit{i)} to create Quality and Feature models, representing the Product-Line Architecture of the ML pipelines, enriched by Quality aspects. This modeling is in charge of ML experts; \textit{ii)} to specify  functional and quality requirements that the final ML system must satisfy. This modeling task is in charge of ML designers.

This paper is organized as follows: in section \ref{sec:rel}, we discuss related works describing first the papers that are related to identify and engineering quality attributes in ML pipelines; next we discuss other approaches that uses Feature Models to model quality of software systems or other kinds of ML pipelines. In section \ref{sec:def}, we make an overview of the problem of Quality Assurance in ML pipelines: we first discuss the selected quality attributes and how they affect ML pipelines, next we describe \our, a more comprehensive approach that aims to define an innovative model-driven
framework that guides ML designers in developing ML pipelines assuring quality requirements. Section \ref{sec:approach} describes the modeling framework that we have developed on this work: we start by describing the Quality and Feature Meta-Model, and next we describe the implemented graphical editor to create Quality and Feature models and specify quality requirements. Section \ref{sec:eval} is dedicated to a proof of concept of the developed modeling framework by reproducing a case study. Finally, section \ref{sec:conclusion} presents some discussions, describes future work, and wraps up the paper.
\section{Related Work}\label{sec:rel}

The problem of quality assurance in machine learning systems has gained much relevance in the last years. Many articles highlight the needing of defining and formalize new standard quality attributes for machine learning systems  \cite{giray2021software,de2019understanding,villamizarrequirements,muccini_software_2021,bosch_engineering_2021,martinez-fernandez_software_2022}. Most of the works in the literature focus either on the identification of the most relevant quality attributes for ML systems or on the formalization of them in the context of pipelines development.

Concerning the identification of quality attributes in ML systems, the authors of  \cite{kumeno2019sofware,zhang2020machine}  identify three main components in which quality attributes can be found: \textbf{Training Data}, \textbf{ML Models} and \textbf{ML Platforms}. The quality of \textbf{Training Data} is usually evaluated with properties such as \textit{privacy}, \textit{bias}, \textit{number of missing values}, \textit{expressiveness}. For \textbf{ML Model}, the authors mean the trained model used by the system. The quality of this component is usually evaluated by \textit{fairness}, \textit{explainability}, \textit{interpretability}, \textit{security}. Finally, the \textbf{ML Platform} is the  implementation of the system, which is affected mostly by \textit{security} and \textit{computational complexity}. 
Muccini et al. identify in \cite{muccini_software_2021} a set of quality properties as stakeholders' constraints and highlight the needing of considering them during the \textit{Architecture Definition} phase. The quality attributes are: \textit{data quality}, \textit{ethics}, \textit{privacy}, \textit{fairness}, \textit{ML models' performance}, etc. Martinez-Fernàndez et al. also highlight in \cite{martinez-fernandez_software_2022} the needing of formalizing quality properties in ML systems and to update the software quality requirements defined by ISO 25000 \cite{iso_isoiec_nodate}. The most relevant properties highlighted by the authors concern: \textit{ML safety}, \textit{ML ethics}, and \textit{ML explainability}. 

Many solutions have been proposed to formalize and model standard quality attributes in ML systems. \textit{CRISP\_ML} is a process model proposed by Studer et al. \cite{studer2021towards}, extending the more known \textit{CRISP\_DL} \cite{martinez2019crisp} process model to machine learning systems. They identify a set of common phases for the building of ML systems namely: \textit{Business and Data understanding}, \textit{Data preparation}, \textit{Modeling}, \textit{Evaluation}, \textit{Deployment}, \textit{Monitoring and Maintenance}. For each phase, the authors identify a set of functional quality properties  to guarantee the quality of such systems.
Similarly, the \textit{Q4AI} consortium proposed a set of guidelines \cite{hamada2020guidelines} for the quality assurance of ML systems for specific domains: \textit{generative systems}, \textit{operational data in process systems}, \textit{voice user interface system}, \textit{autonomous driving} and \textit{AI OCR}. For each domain, the authors identify a set of properties and metrics to ensure quality. Concerning the modelling of quality requirements, Azimi et al. proposed a layered model for the quality assurance of machine learning systems in the context of IoT \cite{azimi2020layered}. The model is made of two layers: \textit{Source Data} and \textit{ML Function/Model}. For the \textit{Source Data}, a set of quality features are defined: \textit{completeness}, \textit{consistency}, \textit{conformity}, \textit{accuracy}, \textit{integrity}, \textit{timeliness}. Machine learning models are instead classified into \textit{predictors}, \textit{estimators} and \textit{adapters} and a set of quality features are defined for each of them: \textit{accuracy}, \textit{correctness}, \textit{completeness}, \textit{effectiveness}, \textit{optimality}. Each system is then influenced by a subset of quality characteristics based on the type of ML model and the required data. 
Ishikawa proposed, instead, a framework for the quality evaluation of an ML system \cite{ishikawa2018concepts}. The framework defines these components for ML applications: \textit{dataset}, \textit{algorithm}, \textit{ML component} and \textit{system}, and, for each of them, proposed an argumentation approach to assess quality. Finally, Siebert et al. \cite{siebert2021construction} proposed a formal modelling definition for quality requirements in ML systems. They start from the process definition in \cite{martinez2019crisp} and build a meta-model for the description of quality requirements. The meta-model is made of the following classes: \textit{Entity} (which can be defined at various levels of abstraction, such as the whole system or a specific component of the system), \textit{Property} (also expressed at different levels of abstraction), \textit{Evaluation} and \textit{Measure} related to the property. Starting from this meta-model, the authors build a tree model to evaluate the quality of the different components of the system. To the best of our knowledge, this is the first attempt at formalizing the quality of ML systems using a model-driven approach.

From this analysis, we can conclude that there is a robust research motivation in formalizing and defining new quality attributes for ML systems. Many attempts have been proposed to solve these issues, and several quality properties, metrics and definitions of ML systems can now be derived from the literature. However few issues are still not fully been addressed: 
\begin{enumerate}
    \item a mapping between quality attributes and components of an ML pipeline;
    \item a modeling of how each quality attribute may influence the development of an ML pipeline;
    \item a modeling of the influence each quality property can have on other attributes;
\end{enumerate}

In this paper, we aim to solve these concerns by proposing a novel model-driven approach which will allow ML designers to model ML pipelines with quality attributes. In particular, we extend the meta-model of the Feature Models to allow the specification of \textit{Quality and Feature Models}. The Quality and Feature model comprises features and quality properties which are related to or can be implemented by the features themselves. In addition, the presented meta-model allows the specification of functional and quality requirements by selecting from the Quality and Feature model only the quality properties that are needed and by specifying attributes related to features.

The proposed approach is similar to the one proposed by Asadi et al. in \cite{asadi_toward_2014}. In their work, the authors present a framework for the automatic configuration of feature models based on non-functional requirements. They start from the extension of feature models to allow the definition of quality attributes associated to features. Then, they use the stakeholders requirements (defined as a set of functional and non-functional requirements with relative constraints among them) and the extended feature model to build a Hierarchical Task Network Planning problem which is finally solved to derive the final configuration that satisfies both functional and non-functional requirements. 
However, in their work the authors consider the traditional quality properties as defined in the ISO 25000. Instead, the new quality properties introduced by ML systems requires a different formulation in terms of modeling. For example, some quality attributes (e.g., fairness) are directly implemented by features in the model which must be selected if the implemented property is required. In our work, we consider these new quality properties introduced by ML systems and make a new modeling of them taking into account their new intrinsic characteristics.
Finally, a similar approach of using Feature Models to model ML pipelines has been used by Di Sipio et al. in \cite{di_sipio_low-code_2021}. In their work the authors use Feature Models to model ML pipelines for Recommender Systems, however they do not consider quality attributes in their approach.

\section{Quality Assurance in ML Pipelines}\label{sec:def}

In this section, we make an overview 
of Quality Assurance (QA) in ML pipelines. In particular, in section \ref{sec:hq} we define \textit{quality} in ML pipelines by selecting the quality properties that are more frequent and relevant in the literature. Next, in section \ref{sec:pipeline} we describe a generic ML pipeline and highlight how the chosen quality properties may influence each phase of the process. Finally, in section \ref{sec:approach} we propose our broaden vision on  engineering ML pipelines with quality constraints leveraging on model-driven generative techniques.
The modeling framework we present in this paper (detailed in Section \ref{sec:artifacts}) is the foundational part of our generative approach and it leverages on the . 
definitions given in these sections. 

\subsection{Considered Quality Attributes}\label{sec:hq}

In software engineering, a Quality Requirement is a requirement that specifies criteria that can be used to quantify or qualify the operation of a system, rather than to specify its behaviors \cite{6365165}. 

To analyze an ML pipeline from a qualitative perspective, we have to determine the Quality Attributes (QA) that we can use to judge the operation of the pipeline, and that can influence the ML designers' decisions. 
To identify the QA to consider, we refer to the literature for ML systems \cite{muccini_software_2021,giray2021software,kumeno2019sofware,villamizarrequirements}. 
\begin{figure}
    \centering
    \includegraphics[width=\linewidth]{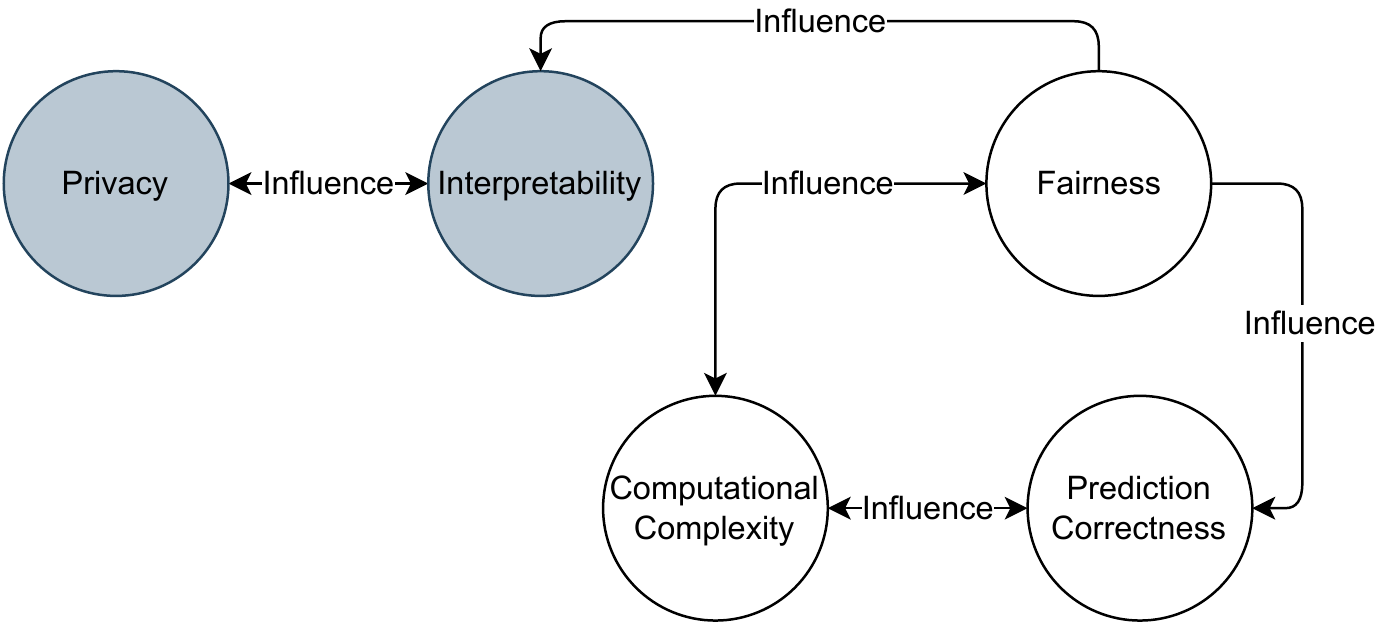}
    \caption{Quality attributes in ML pipelines and their influence}
    \label{fig:qa}
\end{figure}

Figure \ref{fig:qa} summarizes the selected QA and their mutual influence. In the figure, white circles represent the quantitative QA (attributes that can be measured using one or more metrics), while grey ones represent the qualitative QA (attributes that can not be measured with a specific metric). 
Arrows mean influence, for example an arrow from QA $A$ to  QA $B$ models the influences of $A$ to $B$. In some cases, as for the \textit{Computational Complexity} and the \textit{Prediction Correctness}, the impact is mutual; in other cases, like \textit{Fairness} and the \textit{Interpretability}, it is mono-directional. In addition, while some properties can be estimated a-priori (i.e. without adding additional computational tasks to the pipeline) and 
associated with features (e.g., \textit{Interpretability} or \textit{Computational Complexity}), others (e.g., \textit{Fairness} or \textit{Prediction Correctness}) can be evaluated only executing the actual pipeline implementation enriched by specific computational steps that quantifies them. 

In the following, we describe the selected QA in more detail.

\textbf{Computational Complexity.} This quantitative QA defines the computational complexity of the final ML pipeline at production time as the pair \textit{Space Complexity (SC)} and \textit{Time Complexity (TC)}, where the first indicates the amount of memory required by a pipeline to perform all its phases, and the second indicates the time required by the pipeline to complete the whole task \cite{cook1983overview}. This attribute can influence various pipeline stages (see section \ref{sec:pipeline}) and other QA such as \textit{Prediction Correctness}, or \textit{Fairness}. In fact, ML methods that are more accurate in their predictions are also more complex from a computational point of view (e.g., Neural Networks). The same holds for \textit{Fairness}, in fact if the ML pipeline is required to be fair, then enhancing fairness methods must be included in the pipeline and this will increase the overall complexity of the process.

\textbf{Prediction Correctness.} This quantitative QA is used to define how good the model must be in predicting outcomes. There are different metrics in the literature, each addressing a different goal of the user to compute the prediction correctness of an ML model. Among the most common metrics, we cite \textit{Precision}: fraction of true positives (TP) with respect to the total positive predictions \cite{buckland_relationship_1994}; \textit{Recall}: fraction of TP to the total positive items of the dataset \cite{buckland_relationship_1994}; \textit{F1 Score}: harmonic mean of \textit{Precision} and \textit{Recall} \cite{taha_metrics_2015}; \textit{Accuracy}: fraction of True Positives (TP) and True Negatives (TN) above the total of predictions \cite{rosenfield_coefficient_1986}. This QA could impact all the steps of the pipeline regarding the training and testing of an ML model (see section \ref{sec:pipeline}) and can influence the \textit{Computational Complexity} QA.

\textbf{Interpretability.} \textit{Interpretability} can be defined as the ability of a system to enable user-driven explanations of how a model reaches the produced conclusion \cite{carvalho2019machine}. \textit{Interpretability} is one of the quality attributes that can be estimated without executing an actual ML pipeline. Interpretability is a very strong property that can hold only for white-box approaches (such as decision trees). Instead, black-box methods (such as neural networks) requires the addition of explainability enhancing methods to have their results interpretable 
\cite{linardatos2021explainable}. In this paper, we consider interpetability while we leave, for future work, the treatment of explainability for black-box learning approaches.
Interpretability can influence (i.e., it can reduce) the \textit{Privacy} of the pipeline and can be affected by the \textit{Fairness} QA. In fact,  fairness enhancing methods change the value of (some) attributes of the dataset and this can reduce the interpretability of the model \cite{mehrabi_survey_2021}.

\textbf{Privacy.} \textit{Privacy} can be defined as a qualitative QA allowing sensitive information of a dataset to be hidden, changing the value of some attributes \cite{10.1145/1749603.1749605}. Recent works have shown how privacy can affect the ML model during the deployment phase \cite{ruehle_privacy_2021}. This QA directly impacts the \textit{Interpretability}  since a higher level of privacy necessarily causes less interpretability of the model.

\textbf{Fairness.} A ML model can be defined as \textit{fair} if it has no prejudice or favoritism towards an individual or a group based on their inherent or acquired characteristics, identified by the so-called \textit{sensitive variables} \cite{mehrabi_survey_2021}. This quantitative QA influences the \textit{Interpretability} (since some methods for bias mitigation require changing the label of some sensitive attributes), the \textit{Computational Complexity} (since bias mitigation methods add an extra step to the pipeline), and the \textit{Prediction Correctness} (since mitigating the bias typically reduces the performance of the predictions).
\subsection{ML Pipelines with Quality Attributes}\label{sec:pipeline}

"ML pipelines formalize and implement processes to accelerate the development, management, deploy and reuse of ML models" \cite{hapke_building_2020}.
\begin{figure}[h!]
    \centering
    \includegraphics[width=\columnwidth]{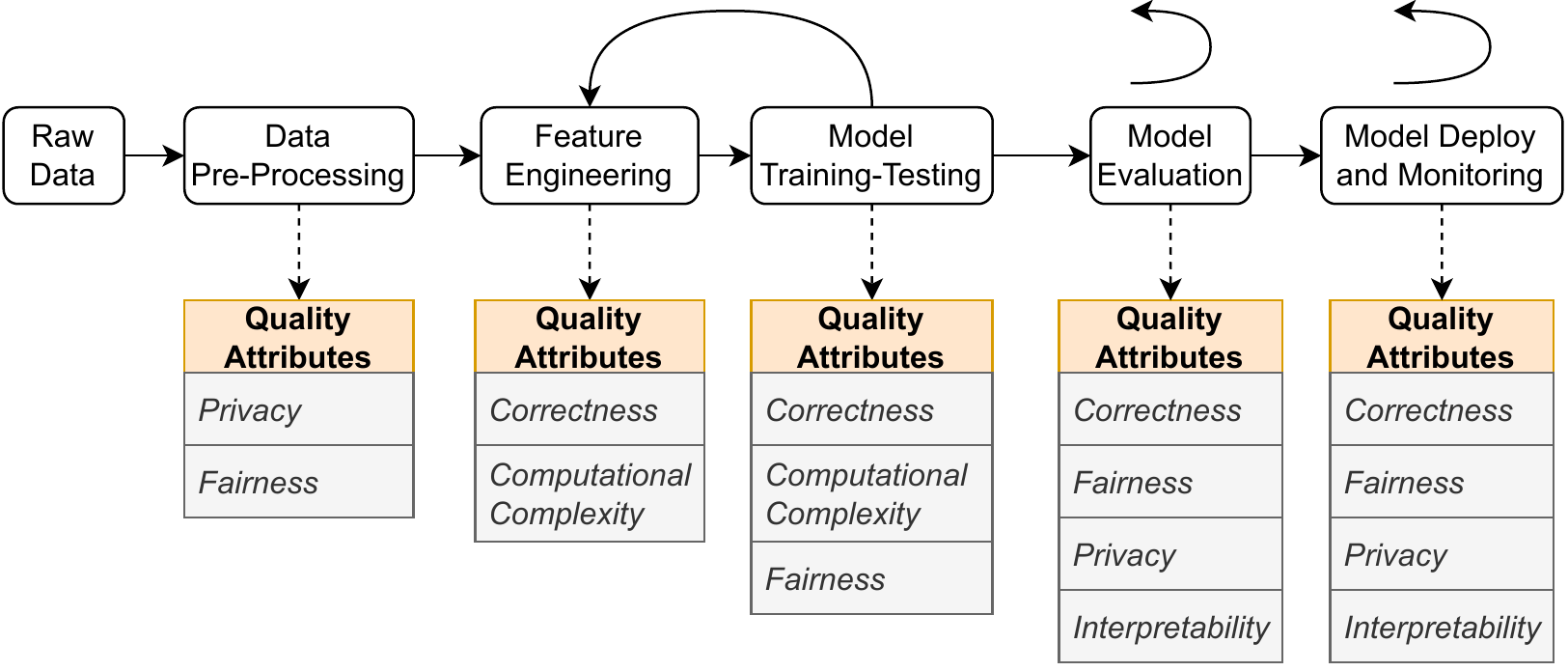}
    \caption{ML pipeline with involved quality attributes (inspired by \cite{amershi2019software,studer2021towards})}
    \label{fig:ds-workflow}
\end{figure}
Figure \ref{fig:ds-workflow}, inspired by \cite{amershi2019software,studer2021towards}, reports a generic ML pipeline (on the top of the figure) together with 
the selected QA affecting the pipeline steps (on the bottom of it). We say that a QA affects a pipeline's step if, in presence of a Quality Requirement (QR) specifying a constraint on that QA, such a QR has an impact on the development of the step or, in other words, it imposes restriction in step's implementation. For instance, in the figure, \textit{Computational Complexity} has an impact on the \textit{Model Training-Testing} and \textit{Feature selection} steps, while \textit{Privacy} can affect \textit{Model Evaluation} and \textit{Model Deploy and Monitoring} steps.

Usually, a ML pipeline takes as input raw (unprocessed) data and, if needed, computes some pre-processing transformations on it. Next, it uses the data to train and test an ML model. This model could be further evaluated, also with human intervention, and finally deployed and continuously monitored. As shown in figure \ref{fig:ds-workflow} the pipeline workflow is an iterative process, meaning that most of the depicted phases can roll back to previous ones. Each of the described steps are affected by at least two of the QA described in section \ref{sec:hq}. 

The \textit{Data Pre-Processing} step, which manipulates data, could be affected by \textit{Privacy} and \textit{Fairness} since they are strongly related to data \cite{10.1145/1749603.1749605,mehrabi_survey_2021}. The \textit{Feature Engineering} step aims at selecting the best dataset's features used to train the ML model. Since this phase is fundamental for the definition of a model able to make good predictions, its implementation is influenced by \textit{Prediction Correctness}. In addition, the \textit{Computational Complexity} is also involved in this phase since the selection of some features can improve or decrease the computational complexity of a model \cite{kearns1990computational}. 


After selecting the best features, the ML model is trained and tested to verify the correctness of the implemented model. In this step, QA affecting its implementation are, besides \textit{Prediction Correctness}, the \textit{Computational Complexity}, and the \textit{Fairness}. In case associated QRs are not satisfied, the ML pipeline rolls back to the \textit{Feature Engineering} to retry the feature selection. 

The last two steps, namely \textit{Model Evaluation} and \textit{Model Deploy and Monitoring}, are affected by the same QAs: \textit{Correctness}, \textit{Fairness}, \textit{Privacy}, and \textit{Interpretability}.


It is worth to note that the described steps are not mandatory and some of them can be skipped if not needed in the specific use case. For instance, if the dataset does not need to be manipulated before training the model, then the \textit{Data Pre-processing} phase is skipped. 

Another crucial aspect to consider when evaluating the quality of ML pipelines is that the depicted QAs are not atomic entities but they can influence each other. 
For example, if the system requires high fairness (e.g., legal reasons \cite{dwork_fairness_2012,kusner_counterfactual_2017,berk_fairness_2018}), 
fairness enhancing components must be included during the \textit{Data Pre-Processing} and the \textit{Model Training and Testing} phases \cite{mehrabi_survey_2021}. However, as explained in section \ref{sec:hq} this has a negative influence on the predictions' correctness \cite{feldman_certifying_2015, kamiran_data_2012}.  So, if the system is required to have also a high \textit{Prediction Correctness}, 
the pipeline development becomes more complicated and the complexity of the problem grows as the number of QAs and QRs increases. 
 
To conclude, the identification and formalization of quality attributes and requirements, and their handling in the development of ML pipelines are mandatory and complex tasks. Our approach takes these challenges and in this paper, as first results, we present the modeling framework that enable the development of quality ML pipelines.

\subsection{The \our Approach}\label{sec:approach}

As explained in section \ref{sec:pipeline}, the quality assurance of ML pipelines is a complex task that must consider numerous variables, such as the impact of quality requirements in each pipeline step and the influence among quality attributes.

The work presented in this paper is part of \our (\ourext), a more comprehensive methodology that aims to define an innovative model-driven approach and relative framework that guides ML experts and designers in developing ML pipelines assuring quality requirements. 
\begin{figure}
    \centering
    \includegraphics[width=\columnwidth]{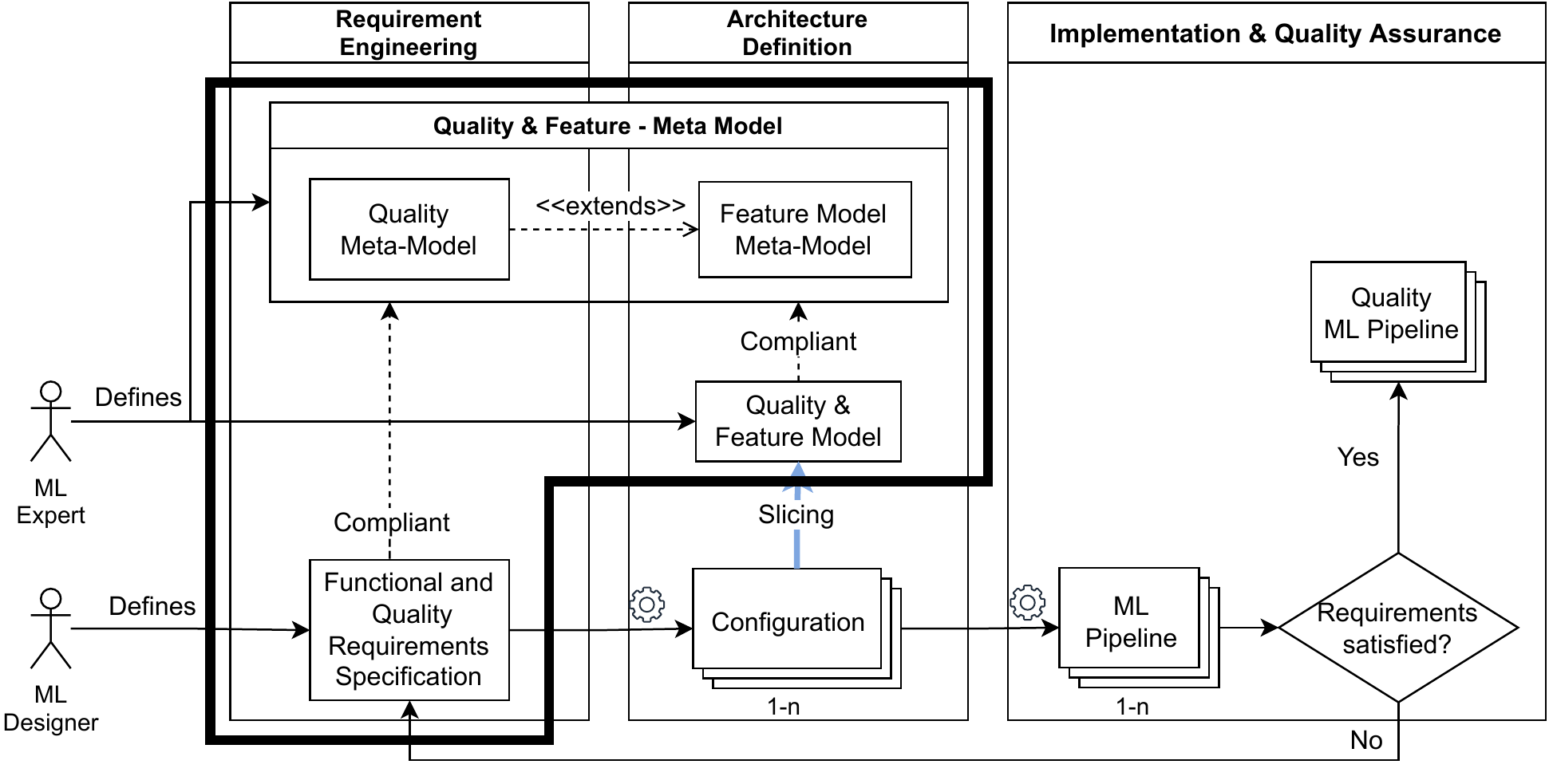}
    \caption{High-level view of \our}
    \label{fig:approach}
\end{figure}
Figure \ref{fig:approach} depicts the high-level architecture of such a framework and highlights, by means of bold line, the modeling framework presented in this paper. 
As discussed earlier (see Figure \ref{fig:ds-workflow}), ML pipelines are made of typical phases \cite{amershi2019software,studer2021towards} that embed a set of standard components identified by the system's functional requirements (like the ML model suited for an ML goal such as classification) and a set of variability points.  These variability points are represented by different methods that implement the functional requirement (e.g., we can have Decision Trees \cite{cramer1976estimation}, Logistic Regression \cite{menard2002applied}, KNN \cite{guo2003knn}, Neural Networks \cite{hagan1997neural}, and other methods that realize classification task). Which one of these will be implemented in the pipeline depends on the specific quality requirements to satisfy. 
For instance, suppose we consider multi-class classification problems \cite{aly2005survey} and fairness quality property. In this case, variability points are represented by ML models suitable for the multi-class classification task and methods to enhance fairness, which must be included in the pipeline if we want to achieve the given quality requirement. However, while some fairness methods can be applied to multi-class datasets (e.g., \textit{Exponentiated Gradient} \cite{agarwal_reductions_2018}, or \textit{Blackbox} \cite{putzel_blackbox_2022}) others can not (e.g., \textit{Reweighing} \cite{kamiran_data_2012}), or \textit{DIR} \cite{feldman_certifying_2015}), hence, they have to be excluded in the pipeline's implementation.

Product-Line Architectures, specified by Feature Models, \cite{thum2011abstract,kang1990feature} 
represent a suitable model to formalize ML pipelines with variability. Feature Models allow us to define a template for families of software products with standard features and a set of variability points that differentiate the final systems. However, they miss adequate means to specify quality attributes and requirements. In fact, they do not allow to specify thresholds, metrics, or quality properties related to features. To address this issue, in \our approach (see Figure \ref{fig:approach}),  we propose to extend the Feature Models meta-model to enable:
\begin{enumerate}
    \item the creation, by the machine learning experts and designers, of an enriched feature model with associated quality attributes for each variability (like done in \cite{asadi_toward_2014});
    \item the specification, by the ML designer, of functional and quality requirements that the implemented pipeline (that is the one with all variability points solved) satisfies.
\end{enumerate}

During the \textit{Requirement Engineering} phase, the end-user (named ML designer in figure \ref{fig:approach}) specifies a set of functional and quality requirements compliant with the defined meta-model. These requirements are used during the \textit{Architecture Definition} step to automatically generate, from the extended feature model provided by the machine learning expert, a set of ML pipeline configurations able to satisfy the defined functional requirements (the \textit{Configuration} boxes in the figure). The configurations are defined by removing from the feature model all the components (and their relative specification) not suitable to meet the specified requirements or against some constraints specified in the model. The generated configurations are given as input to the \textit{Implementation \& Quality Assurance} step that aims to:
\begin{enumerate}
    \item generate for each configuration a python script implementing the \textit{ML Pipeline};
    \item verify that at least one generated pipeline satisfies the quality requirements (\textit{Quality ML Pipeline}).
\end{enumerate}
The framework returns the set of Quality ML Pipelines, satisfying quality constraints, if any, or demands the ML designer to relax quality requirements and repeat the process.

As highlighted in figure \ref{fig:approach}, the work presented in this paper implements the \textit{Requirement Engineering} and part of the \textit{Architecture Definition} phases. In particular, we extend the Feature Models meta-model with the possibility of associating quality properties to features and specifying functional and quality requirements. To enact the modeling, we developed on top of the extended meta-model. 
In addition, in section \ref{sec:eval}, we show the potentialities of \our modeling approach. 

\section{MANILA Modeling Framework}\label{sec:artifacts}

This section describes the \our modeling framework which is made of two main components: \textit{i)} the Quality and Feature meta-model, and \textit{ii)} a graphical editor to create the quality and feature models. Using them, it is possible to specify:
\begin{itemize}
    \item a feature model for each ML problem. Each feature model represents the ML pipeline for the specific ML problem, where the variability points are the set of alternative methods that could be used in the pipeline steps. This modeling action is done by the ML expert once. The produced models are stored in a repository for future uses of ML designers or for models updates from the experts;
    \item the quality characteristics, possessed by the methods modelled as variability points. This task is still in charge of the ML expert and can be specified only for QA quantifiable a-priori;
    \item the functional and quality requirements of the specific ML problem to solve. This modeling task is a duty of ML designer and it is repeated every time a new ML system must be implemented.
\end{itemize}

All the artifacts, models and meta-models, are available at the following link:
\url{https://github.com/giordanoDaloisio/manila-framework}. 

In the following, we detail the \textit{Quality and Feature meta-model} and the editor we implement to perform modeling tasks, while an example of modeling is showed, as proof of concept, in section \ref{sec:eval}.

\begin{figure*}
    \centering
    \includegraphics[width=\textwidth]{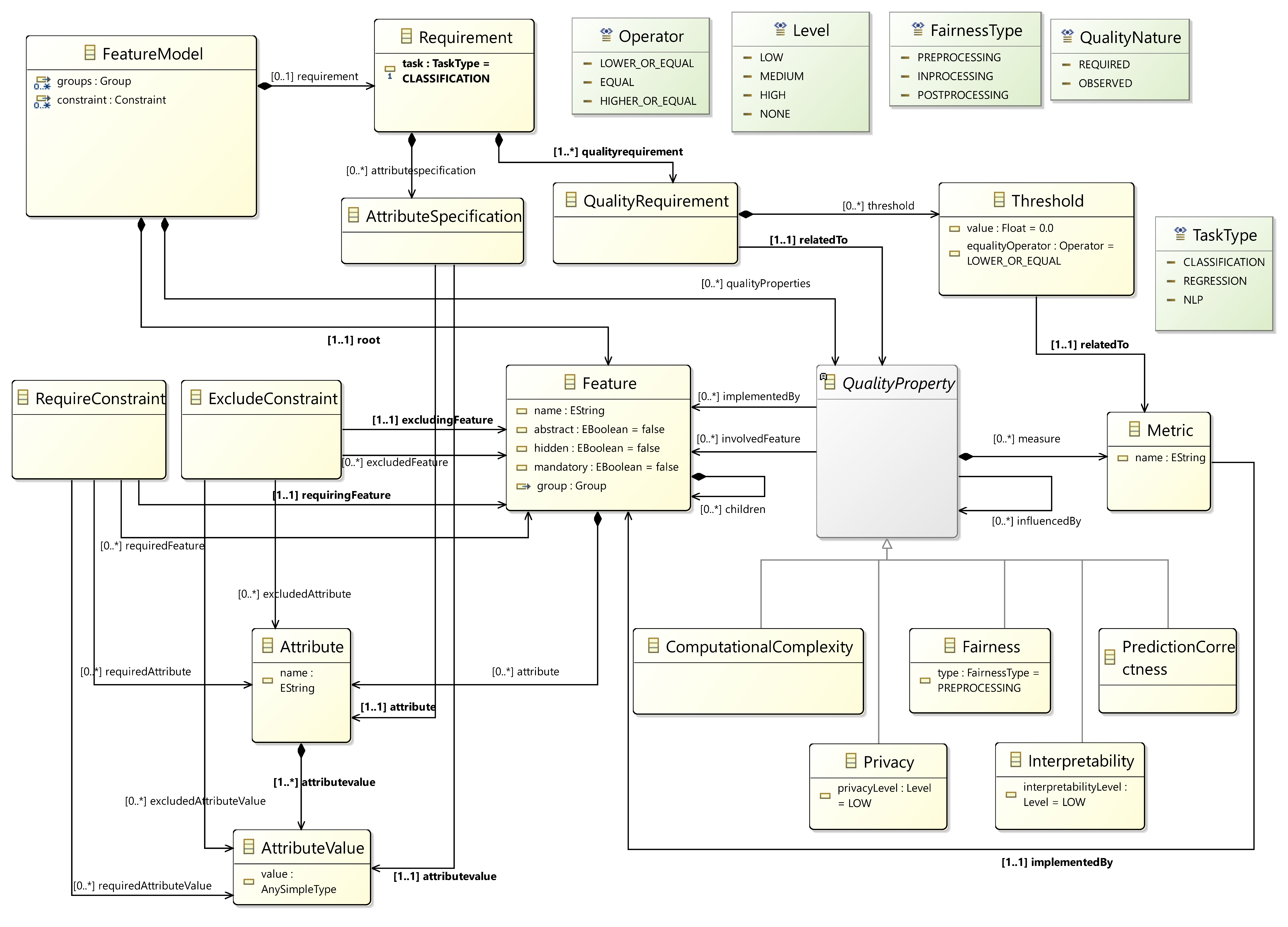}
    \caption{Quality and Feature meta-model}
    \label{fig:meta-model}
\end{figure*}

\subsection{Quality and Feature Meta-Model}\label{sec:meta}

The \textit{Quality and Feature meta-model} allows the definition of an extended Feature Model with quality properties and the specification of functional and quality requirements. The QAs identified in section \ref{sec:hq} are used to specify both quality properties of features and quality requirements of the whole pipeline. Quality properties and requirements refer to QAs and show the same form, but the former models \textit{descriptive characteristics} of variability points whereas the latter models the \textit{prescriptive characteristic} of the whole ML pipeline. The combination of features, showing specific quality characteristics can (or not) satisfy the quality requirements of the ML pipeline. \our approach aims to select a combination of features that all together implement the ML pipeline and satisfy the quality requirements.

The meta-model has been implemented using the established EMF Framework \cite{steinberg_emf_2008} and is depicted in figure \ref{fig:meta-model}.

As already discussed in section \ref{sec:approach}, Feature Models are a suitable model to formalize ML pipelines where variability points models the concrete methods that can use to implement a pipeline's step. Hence, \our meta-model is an extension of the meta-model proposed in \cite{burdek_reasoning_2016}.

The root entity of the meta-model is the \textit{Feature Model}, which represents the model itself and is composed by a \textit{Feature} entity, representing the whole ML pipeline, and composed itself by a set of children, i.e. (sub-)\texttt{Features}. 

A \textit{Feature} entity represents a building block of the ML pipeline. Features have a name and it can be abstract (i.e., they do not have a concrete implementation in the final system), hidden (i.e., they are not shown in the model) and mandatory (i.e., they must be selected in every configuration) \cite{kang1990feature}. If a feature does not show the abstract attribute, it means that it will be concrete. 
A \textit{Feature} can be a composition of other Features. Variability points are represented by the \textit{Feature}s as well, but at a lower level in the composition. 

In the composition, Features can belong to two groups: \textit{OrGroup} and \textit{AltGroup}\footnote{To enhance the visualization, we omitted from figure \ref{fig:meta-model} these entities. However, a full picture of the meta-model is available here \url{https://github.com/giordanoDaloisio/manila-framework/blob/main/assets/metamodel.png}}. \textit{OrGroup} entities represent an inclusive relationship, meaning that at least one child must be selected if the father is selected. \textit{AltGroup} entities represent instead an exclusive relationship, meaning that exactly one child must be selected if the father is selected. 

Finally, Features can have \textit{Attributes}. Each \textit{Attribute} has a name and contains a set of \textit{Attribute Value}, predefined by the ML expert, that represent the set of possible values an \textit{Attribute} can have. 

\textit{Feature}, \textit{Attribute}, and \textit{AttributeValue} entities can be part of a logical \textit{Constraint}. There are two types of logic constraints: \textit{RequireConstraint} and \textit{ExcludeConstraint}. The first means that, if a feature involved in the RequiredConstraint is selected in a configuration, then also the other Features or Attributes in the logical constraint must be considered in the final implementation.  The second means instead that, if a feature  under the ExcludeContraint is selected, then the other entity must not be selected in the configuration. 

Note that, we extended the original Feature Meta-model to include in the Constraints also Attributes and AttributeValues. This extension can help to automatically select suitable components starting from, for example, the characteristic of the Dataset, which might have several attributes (e.g. the number of classes in the case of a classification tasks or the number of sensitive variables when dealing with fairness \cite{mehrabi_survey_2021}) whose values are not compatible with the characteristics of components that hence must be excluded during the generation of the ML pipeline configuration.

A Feature Model can contain a set of \textit{QualityProperty}. Following the definition of quality we made in section \ref{sec:hq},
each quality property refers to one of the attributes we defined earlier: \textit{Computational Complexity}, \textit{Privacy}, \textit{Fairness}, \textit{Interpretability}, and \textit{Prediction Correctness}. Measurable quality properties are composed by one or more \textit{Metric} entities, each implemented by a feature.

\textit{Quality property} itself can be \textit{implemented by} one or more Features and it can involve other Features. The distinction between \textit{implemented by} and \textit{involved feature} is needed since some features directly implement some quality properties (e.g., fairness is straight provided by some methods able to mitigate the bias of an ML method), otherwise others are an intrinsic quality property without requiring an extra computational step (e.g., the computational complexity is an intrinsic feature of ML algorithms). Finally, quality properties can also be influenced by each other (modelled as \textit{influenced by} relation). 

Finally, the functional and quality requirements are represented by the \textit{Requirement} entity. In particular, a Feature Model can include a \textit{Requirement} related to a specific ML \textit{task} (classification, regression, natural language processing, and so on). Each \textit{Requirement} is composed by a set of \textit{Attribute Specification}, modeling the functional requirement,  and a set of \textit{Quality Requirement}, modeling the quality requirements on the ML \textit{task}. Each \textit{Attribute Specification} is related to an \textit{Attribute} and an \textit{Attribute Value}. Each \textit{Quality Requirement} is associated with a \textit{Quality Property} previously defined in the model and can contain one or more \textit{Thresholds} related to a specific \textit{Metric}.

\subsection{Graphical Editor}

After defining the meta-model, we implemented a graphical editor that allows users to easily model ML pipelines and specify functional and quality requirements. The editor has been implemented using \textit{Sirius}\cite{viyovic_sirius_2014}, a well-known tool for building graphical editors based on the EMF framework. 

\begin{figure}
    \centering
    \includegraphics[width=\linewidth]{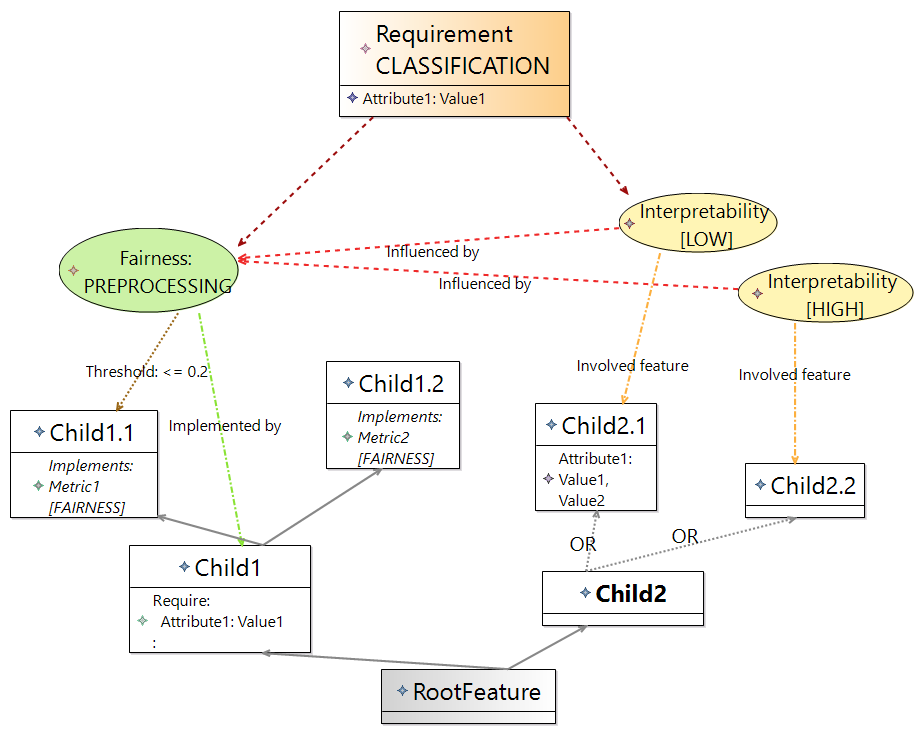}
    \caption{Example of Quality and Feature model with a given requirement}
    \label{fig:example}
\end{figure}

Figure \ref{fig:example} shows an example of a model complaint to the \textit{Quality and Feature Meta-model} built with our editor. Note that, in order to keep the model simpler and focus only on its semantic, we named each feature with abstract label such as \textit{Child} that must be interpreted as components. 
In the figure, \textit{Child2.1} and \textit{Child2.2} represent specific ML pipeline components involved in the \textit{Interpretability} quality property, whereas \textit{Child1} and his children model additional computational components implementing fairness. A real example of modeling can be seen in section \ref{sec:eval}.

In the devised editor, features are represented by boxes: white boxes model concrete features, while grey boxes represent abstract features  \cite{thum2011abstract}. Mandatory features are instead highlighted with a boldface name. Attributes of a feature are defined inside the box, below its name, with a label following the pattern: \texttt{<Attribute Name>: <Attribute Values>}. Concerning the example in figure \ref{fig:example}, \textit{Root Feature} is an abstract feature while all the others are concrete ones. Furthermore, \textit{Child2} is mandatory, and \textit{Child2.1} has an attribute named \textit{Attribute1} with two possible values: \textit{Value1} and \textit{Value2}.

Following the original definition of Feature Models, each feature can have one or more children, and children can belong to logical \textit{OrGroup}  or \textit{AltGroup} \cite{kang1990feature}. In the graphical editor, this information is represented  using dotted edges labeled with \texttt{OR} or \texttt{ALT} keyword. If a child feature does not belong to any group, the edge is not labelled.

In addition, as already said in section \ref{sec:meta}, features can also require or exclude other features or attributes. This constraint is expressed using, inside the box of the requiring/excluding feature, a specification following the form: \texttt{<Require|Exclude>: <Feature Name|Attribute Name|Attribute Name: Attribute Values>}. In the example shown in figure \ref{fig:example}, \textit{Child2.1} and \textit{Child2.2} belong to an \textit{OR} group, while \textit{Child1}, if selected, requires \textit{Attribute1} equal to \textit{Value1}.

QAs are represented with ellipses distinguished by color according to the property type (i.e., fairness is green, interpretability is yellow, and so on). 

Quality Properties (ellipses) are linked to the feature that implement them using a green edge. The editor use a yellow edge to link Quality Properties (ellipses) to the involved features. To keep the model clearer, we have decided not to represent metrics but just to label the features implementing a given metric with a label following the pattern: \texttt{Implements: <Metric Name> [<Quality Property>]}. 

The influence among quality properties is modelled by means of a red edge (\textit{inflienced by}) that starts from the influenced quality (ellipse) and points to the influencing property (ellipse). In the example of figure \ref{fig:example}, we have two quality properties: fairness (of \textit{PREPROCESSING} type \cite{mehrabi_survey_2021}) and interpretability. Fairness requires extra computational task is implemented by the feature \textit{Child1}, while interpretability, being an a-priori quality does not require an extra feature and involves \textit{Child2.1} and \textit{Child2.2}. In particular, \textit{Child2.1} is involved with a low level of interpretability, while \textit{Child2.2} is involved with a high level of interpretability. Interpretability could be influenced by fairness (as discussed in section \ref{sec:hq}). This information is modelled by means two red edges starting from each \textit{Interpretability} node and pointing to the \textit{Fairness} node. Finally, \textit{Child1.1} and \textit{Child1.2} are two features implementing two metrics for fairness, namely \textit{Metric1} and \textit{Metric2}, respectively.

Requirements are represented with orange boxes. The name of the box follows this pattern: \texttt{Requirement <ML Task>}. In the Requirement box, there is a list of possible attribute specifications. Each specification follows the pattern: \texttt{<Attribute Name>: <Attribute Value>}. Quality requirements are represented with red edges connecting the requirement to the related quality property. Finally, thresholds are represented with brown edges that connect the quality property with a feature implementing the relative metric. The value of the threshold is reported with a label on the edge. Concerning figure \ref{fig:example}, there is a requirement for the classification task, setting \textit{Attribute1} equal to \textit{Value1} and requiring both fairness and interpretability quality properties. In particular, there is a requirement for a low level of interpretability and a fairness value equal to or lower than 0.2 under \textit{Metric1}.
\section{Proof of concept}\label{sec:eval}

\begin{figure*}[t!]
    \centering
    \includegraphics[width=\textwidth]{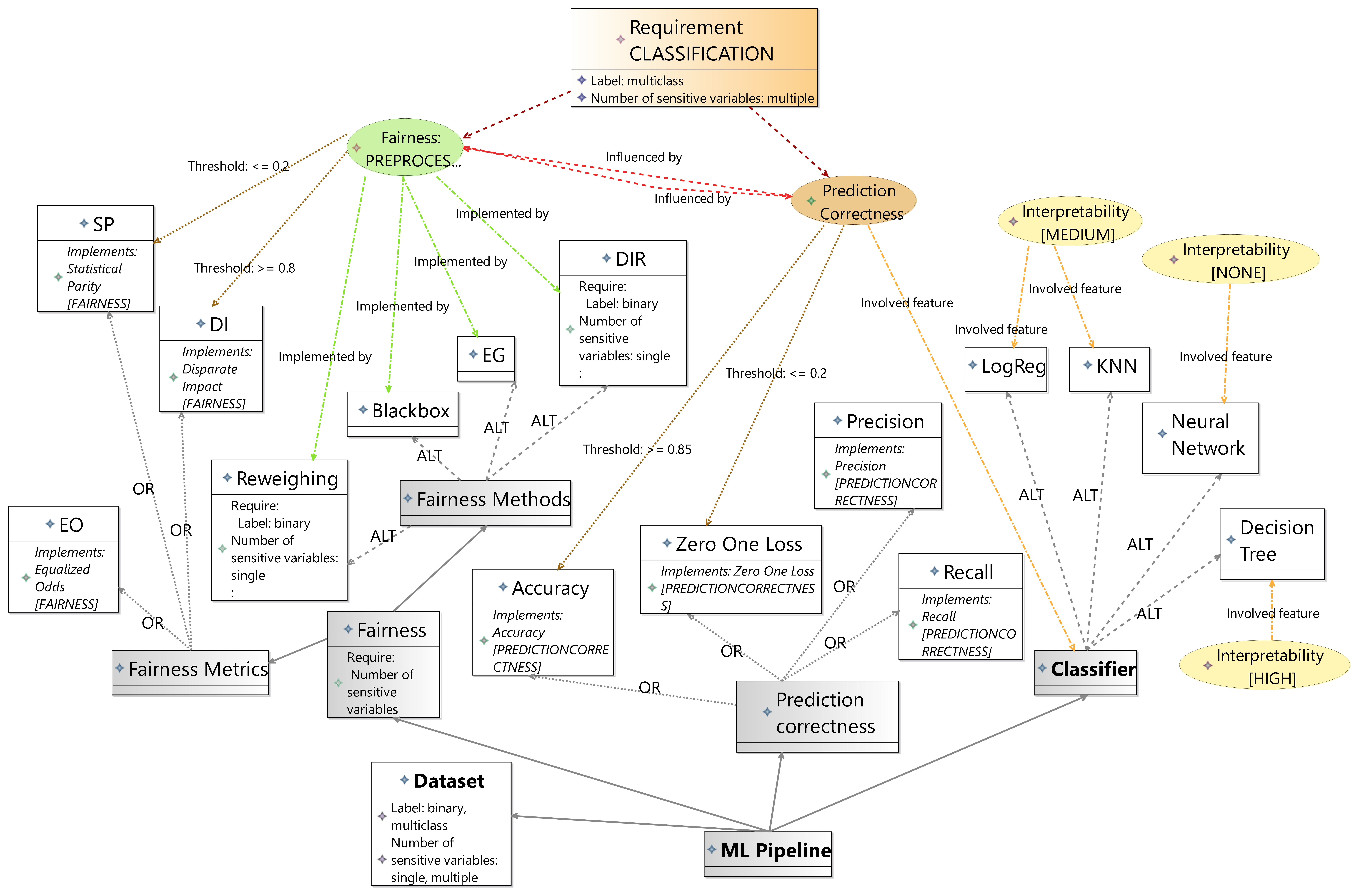}
    \caption{Implemented Quality and Feature model for multi-class classification task with given requirements}
    \label{fig:quality_spec}
\end{figure*}

This section describes the proof of concept we conducted in our work. In particular, we want to ask the following research questions: 

\begin{tcolorbox}
    \begin{questions}
    \item Are our meta-model and editor able to properly represent an actual ML pipeline specification with involved quality requirements?\label{rq1} 
    \item Does the built model enable the generation of ML pipeline configurations?
    \label{rq2}
    \end{questions}
\end{tcolorbox}


To answer these questions, we specify, using the implemented editor, a Quality and Feature model that reproduces a ML pipeline for multi-class classification problems, subject to fairness and prediction correctness quality constraints. In particular, in this example we focus on the Pre-Processing fairness, which consists on applying fairness enhancing methods to the dataset before using it as an input to train the classifier \cite{mehrabi_survey_2021}. Hence, recalling the figure \ref{fig:ds-workflow}, the pipeline steps involved in this example are \textit{Data Pre-Processing} and \textit{Model Training-Testing}. We do not represent the \textit{Feature Engineering} phase to keep model clearer and more readable.

Concerning \ref{rq1}, figure \ref{fig:quality_spec} shows the implemented model to represent the multi-class classification pipeline\footnote{The same picture with a higher resolution is available here \url{https://github.com/giordanoDaloisio/manila-framework/blob/main/assets/quality-features-model.jpg}}. 
The root feature is the \textit{ML Pipeline}, which is an abstract mandatory feature. Describing  its children, going from left to right, the first is the \textit{Dataset} feature, which has two attributes regarding the type of label (binary or multiclass) and the number of sensitive variables (i.e., variables affected by bias). Following, there are the features related to fairness. In particular, these features are distinguished between methods to implement fairness and metrics to measure it. Concerning methods, we have included the most common pre-processing methods to improve fairness: \textit{Reweighing} \cite{kamiran_data_2012}, \textit{DIR} \cite{feldman_certifying_2015}, \textit{EG} \cite{agarwal_reductions_2018}, and \textit{Blackbox} \cite{putzel_blackbox_2022}. Since \textit{Reweighing} and \textit{DIR} can only be applied to datasets with a binary label and a single sensitive variable, we added a \textit{Require} logical constraint requiring the \textit{Label} attribute to be equal to \textit{binary} and the \textit{Number of sensitive variables} attribute to be equal to \textit{single}. In addition, the abstract \textit{Fairness} feature requires, if selected, to specify the \textit{Number of sensitive variables} attribute. 

The \textit{Fainess Metrics} entity is in relationship with features implementing metrics usable to measure fairness, which are: \textit{Statistical Parity (SP)} \cite{kusner_counterfactual_2017}, \textit{Disparate Impact (DI)} \cite{dwork_fairness_2012}, and \textit{Equalized Odds (EO)} \cite{hardt_equality_2016}. These features belong to an \textit{OR} group, meaning that at least one metric must be selected if the father is selected.

Features implementing the metrics for the \textit{Prediction Correctness} are \textit{Precision} \cite{buckland_relationship_1994}, \textit{Recall} \cite{buckland_relationship_1994}, \textit{Zero One Loss} \cite{domingos_optimality_1997}, and \textit{Accuracy} \cite{rosenfield_coefficient_1986}. These features are children of an abstract father feature named \textit{Prediction correctness} and belong to an \textit{OR} group.

Finally, the model contains features representing ML classifiers. In this model, we have included the most used ML methods for classification: \textit{KNN} \cite{guo2003knn}, \textit{Logistic Regression} \cite{menard2002applied}, \textit{Neural Networks} \cite{hagan1997neural}, and \textit{Decision Trees} \cite{cramer1976estimation}. These features are children of an abstract \textit{Classifier} feature, which is mandatory, and they are members of an \textit{ALT} group, meaning that only one of them can be selected inside a configuration. The  \textit{Classifier}  is involved in the \textit{Prediction Correctness} quality property, meaning that all the children of this feature will be also involved in this quality attribute. Fairness and prediction correctness can influence each other, so there are two (red) edges indicating their mutual influence.

The presented model specifies the requirements of the needed ML system that must execute multi-class classification on datasets with more than one sensitive variable. The ML system must be fair and have high prediction correctness:

\begin{enumerate}
    \item the Prediction correctness must consider \textit{Accuracy} metric, whose value must be higher or equal to $0.85$,  and \textit{Zero-One Loss} metric that must be lower or equal to $0.2$;
    \item Fairness must use \textit{Disparate Impact} metrics, whose value must be higher or equal to $0.8$, and \textit{Statistical Parity} metrics that must be lower or equal to $0.2$.
\end{enumerate}

These functional and quality requirements are modeled through a \textit{Requirement} entity, which specifies in the \textit{Dataset} feature, the value \textit{multi-class} for the attribute \textit{Label} and the value \textit{multiple} for the attribute \textit{Number of sensitive variables}. 

The \textit{Requirement} is connected to the two involved quality properties: \textit{Prediction correctness} and \textit{Fairness}. The two involved properties are then related to the features implementing the metrics used in the thresholds. In particular, \textit{Prediction Correctness} is connected to \textit{Accuracy} and \textit{Zero One Loss} through edges labeled with the threshold's value. Same for \textit{Fairness}, which is connected to \textit{SP} and \textit{DI}. 

The model also specifies the \textit{Interpretability} property that is particular important for classification problems even if it is not targeted by the showed requirement. But, in general, another ML system specification could ask for it. 

\begin{tcolorbox}
    \ref{rq1}. The presented meta-model is able to represent an ML pipeline with involved quality properties. In particular, relying on the concept of feature models, we can represent any step of an ML pipeline using features and constraints. Each feature can be \textit{involved in} or \textit{can implement} one or more quality properties. These properties, along with features' attributes, are then used to specify a requirement. The requirement, along with the constraints defined in the model, can be used to generate a set of ML pipeline configurations by cutting off the features that are not needed and those against a particular constraint.
\end{tcolorbox}

\begin{table*}[ht!]
\centering
\caption{ML pipeline configurations derived from the requirement specification}
\label{tab:configs}
\resizebox{\textwidth}{!}{%
\begin{tabular}{|l|c|c|c|c|c|c|c|c|c|c|c|c|c|c|c|c|}
\hline
 & \multicolumn{1}{|l|}{} & \multicolumn{4}{|c|}{\textbf{Classifier}} & \multicolumn{4}{|c|}{\textbf{Prediction Correctness Metrics}} & \multicolumn{4}{|c|}{\textbf{Fairness Methods}} & \multicolumn{3}{|c|}{\textbf{Fairness Metrics}} \\ \hline
\multicolumn{1}{|c|}{} & \textbf{Dataset} & \textbf{KNN} & \textbf{LogReg} & \textbf{Neural Net} & \textbf{Decision Tree} & \textbf{Zero One Loss} & \textbf{Accuracy} & \textbf{Precision} & \textbf{Recall} & \textbf{Blackbox} & \textbf{EG} & \textbf{Reweighing} & \textbf{DIR} & \textbf{DI} & \textbf{SP} & \textbf{EO} \\ \hline
1 & x & x &  &  &  & x & x &  &  & x &  &  &  & x & x &  \\ \hline
2 & x & x &  &  &  & x & x &  &  &  & x &  &  & x & x &  \\ \hline
3 & x &  & x &  &  & x & x &  &  & x &  &  &  & x & x &  \\ \hline
4 & x &  & x &  &  & x & x &  &  &  & x &  &  & x & x &  \\ \hline
5 & x &  &  & x &  & x & x &  &  & x &  &  &  & x & x &  \\ \hline
6 & x &  &  & x &  & x & x &  &  &  & x &  &  & x & x &  \\ \hline
7 & x &  &  &  & x & x & x &  &  & x &  &  &  & x & x &  \\ \hline
8 & x &  &  &  & x & x & x &  &  &  & x &  &  & x & x &  \\ \hline
\end{tabular}%
}
\end{table*}

Concerning \ref{rq2}, the requirement of our example defines a multi-class dataset with more than one sensitive variable. This specification automatically excludes \textit{Reweighing} and \textit{DIR} from the features to consider to improve fairness since they require a dataset with a binary label and one sensitive variable. In addition, \textit{Precision}, \textit{Recall}, and \textit{Equalized Odds} are metrics not considered in the given quality specification, so their implementations can be omitted from the set of configurations. Finally, concerning the classifier methods, they are not involved in any specification or constraint, so all of them must be considered. However, since they belong to an \textit{ALT} group, they must be included in different configurations. Table \ref{tab:configs} summarises the set of valid configurations that can be defined from this requirement. In particular, each row of the table depicts a valid ML pipeline configuration in which only the required features are selected. It can be noted how some features (like the dataset or the fairness and correctness metrics) are always selected. Other features, like the different ML classifiers or the fairness enhancing methods varies between configurations. Finally, there are features that are not present in any configuration, because they are against the functional and quality requirements defined above.

\begin{tcolorbox}
    \ref{rq2}. Underlying logical conditions can be derived from the requirement and constraints specified by the designed model. The logical condition automatically defines a set of pipeline configurations by cutting off the features that are not needed.
\end{tcolorbox}

\section{Conclusion and Future Work}\label{sec:conclusion}

In this paper, we have presented a novel approach to model and specify ML pipelines with quality requirements. First, we have identified the most influential quality properties in ML pipelines by selecting the quality attributes that are most cited in the literature. We have shown how these quality properties may impact the pipeline's steps. Next, we have presented \our, a novel model-driven approach that will guide ML designers in developing ML pipelines assuring quality requirements. Finally, we described the modeling framework implemented in our work and proven its usefulness by reproducing a real use case scenario.

In particular, in order to model ML pipelines with quality attributes, we relied on the concept of Feature Models. We extended the Feature Models meta-model to create a Quality and Features meta-model, which allows associating quality attributes to features and specifying functional and quality requirements. We have demonstrated the expressiveness of our meta-model by reproducing an ML pipeline with quality attributes. In addition, we implemented a graphical editor which allows the creation of Quality and Feature models and the specification of the requirements.

In the future, we plan first to conduct a user evaluation of our graphical editor to evaluate its usability and to integrate other quality properties (for instance, \textit{Explainability} \cite{linardatos2021explainable}) in our meta-model. Next, we will continue the development of \our framework by first transforming the derived ML pipeline configurations into actual implementations and further by evaluating the real quality of the implemented pipelines through their execution in a test environment. This last step will allow the ML designer to select the most promising ML system among the ones that satisfy all the requirements.

\vspace{0.5em}

\noindent\textbf{Acknowledgments.} \TaSoBdAck

\bibliographystyle{ACM-Reference-Format}
\bibliography{bibliography}


\begin{thebibliography}{54}


\ifx \showCODEN    \undefined \def \showCODEN     #1{\unskip}     \fi
\ifx \showDOI      \undefined \def \showDOI       #1{#1}\fi
\ifx \showISBNx    \undefined \def \showISBNx     #1{\unskip}     \fi
\ifx \showISBNxiii \undefined \def \showISBNxiii  #1{\unskip}     \fi
\ifx \showISSN     \undefined \def \showISSN      #1{\unskip}     \fi
\ifx \showLCCN     \undefined \def \showLCCN      #1{\unskip}     \fi
\ifx \shownote     \undefined \def \shownote      #1{#1}          \fi
\ifx \showarticletitle \undefined \def \showarticletitle #1{#1}   \fi
\ifx \showURL      \undefined \def \showURL       {\relax}        \fi
\providecommand\bibfield[2]{#2}
\providecommand\bibinfo[2]{#2}
\providecommand\natexlab[1]{#1}
\providecommand\showeprint[2][]{arXiv:#2}

\bibitem[\protect\citeauthoryear{Agarwal, Beygelzimer, Dudik, Langford, and
  Wallach}{Agarwal et~al\mbox{.}}{2018}]%
        {agarwal_reductions_2018}
\bibfield{author}{\bibinfo{person}{Alekh Agarwal}, \bibinfo{person}{Alina
  Beygelzimer}, \bibinfo{person}{Miroslav Dudik}, \bibinfo{person}{John
  Langford}, {and} \bibinfo{person}{Hanna Wallach}.}
  \bibinfo{year}{2018}\natexlab{}.
\newblock \showarticletitle{A {Reductions} {Approach} to {Fair}
  {Classification}}. In \bibinfo{booktitle}{\emph{Proceedings of the 35th
  {International} {Conference} on {Machine} {Learning}}}.
  \bibinfo{publisher}{PMLR}, \bibinfo{pages}{60--69}.
\newblock
\urldef\tempurl%
\url{https://proceedings.mlr.press/v80/agarwal18a.html}
\showURL{%
\tempurl}
\newblock
\shownote{ISSN: 2640-3498.}


\bibitem[\protect\citeauthoryear{Aly}{Aly}{2005}]%
        {aly2005survey}
\bibfield{author}{\bibinfo{person}{Mohamed Aly}.}
  \bibinfo{year}{2005}\natexlab{}.
\newblock \showarticletitle{Survey on multiclass classification methods}.
\newblock \bibinfo{journal}{\emph{Neural Netw}} \bibinfo{volume}{19},
  \bibinfo{number}{1-9} (\bibinfo{year}{2005}), \bibinfo{pages}{2}.
\newblock


\bibitem[\protect\citeauthoryear{Amershi, Begel, Bird, DeLine, Gall, Kamar,
  Nagappan, Nushi, and Zimmermann}{Amershi et~al\mbox{.}}{2019}]%
        {amershi2019software}
\bibfield{author}{\bibinfo{person}{Saleema Amershi}, \bibinfo{person}{Andrew
  Begel}, \bibinfo{person}{Christian Bird}, \bibinfo{person}{Robert DeLine},
  \bibinfo{person}{Harald Gall}, \bibinfo{person}{Ece Kamar},
  \bibinfo{person}{Nachiappan Nagappan}, \bibinfo{person}{Besmira Nushi}, {and}
  \bibinfo{person}{Thomas Zimmermann}.} \bibinfo{year}{2019}\natexlab{}.
\newblock \showarticletitle{Software engineering for machine learning: A case
  study}. In \bibinfo{booktitle}{\emph{2019 IEEE/ACM 41st International
  Conference on Software Engineering: Software Engineering in Practice
  (ICSE-SEIP)}}. IEEE, \bibinfo{pages}{291--300}.
\newblock


\bibitem[\protect\citeauthoryear{Asadi, Soltani, Gasevic, Hatala, and
  Bagheri}{Asadi et~al\mbox{.}}{2014}]%
        {asadi_toward_2014}
\bibfield{author}{\bibinfo{person}{Mohsen Asadi}, \bibinfo{person}{Samaneh
  Soltani}, \bibinfo{person}{Dragan Gasevic}, \bibinfo{person}{Marek Hatala},
  {and} \bibinfo{person}{Ebrahim Bagheri}.} \bibinfo{year}{2014}\natexlab{}.
\newblock \showarticletitle{Toward automated feature model configuration with
  optimizing non-functional requirements}.
\newblock \bibinfo{journal}{\emph{Information and Software Technology}}
  \bibinfo{volume}{56}, \bibinfo{number}{9} (\bibinfo{date}{Sept.}
  \bibinfo{year}{2014}), \bibinfo{pages}{1144--1165}.
\newblock
\showISSN{0950-5849}
\urldef\tempurl%
\url{https://doi.org/10.1016/j.infsof.2014.03.005}
\showDOI{\tempurl}


\bibitem[\protect\citeauthoryear{Azimi and Pahl}{Azimi and Pahl}{2020}]%
        {azimi2020layered}
\bibfield{author}{\bibinfo{person}{Shelernaz Azimi} {and}
  \bibinfo{person}{Claus Pahl}.} \bibinfo{year}{2020}\natexlab{}.
\newblock \showarticletitle{A Layered Quality Framework for Machine
  Learning-driven Data and Information Models.}. In
  \bibinfo{booktitle}{\emph{ICEIS (1)}}. \bibinfo{pages}{579--587}.
\newblock


\bibitem[\protect\citeauthoryear{Berk, Heidari, Jabbari, Kearns, and Roth}{Berk
  et~al\mbox{.}}{2018}]%
        {berk_fairness_2018}
\bibfield{author}{\bibinfo{person}{Richard Berk}, \bibinfo{person}{Hoda
  Heidari}, \bibinfo{person}{Shahin Jabbari}, \bibinfo{person}{Michael Kearns},
  {and} \bibinfo{person}{Aaron Roth}.} \bibinfo{year}{2018}\natexlab{}.
\newblock \showarticletitle{Fairness in Criminal Justice Risk Assessments: The
  State of the Art}.
\newblock \bibinfo{journal}{\emph{https://doi.org/10.1177/0049124118782533}}
  \bibinfo{volume}{50}, \bibinfo{number}{1} (\bibinfo{year}{2018}),
  \bibinfo{pages}{3--44}.
\newblock
\showISSN{15528294}
\urldef\tempurl%
\url{https://doi.org/10.1177/0049124118782533}
\showDOI{\tempurl}
\newblock
\shownote{Publisher: {SAGE} {PublicationsSage} {CA}: Los Angeles, {CA}.}


\bibitem[\protect\citeauthoryear{Bosch, Olsson, and Crnkovic}{Bosch
  et~al\mbox{.}}{2021}]%
        {bosch_engineering_2021}
\bibfield{author}{\bibinfo{person}{Jan Bosch},
  \bibinfo{person}{Helena~Holmström Olsson}, {and} \bibinfo{person}{Ivica
  Crnkovic}.} \bibinfo{year}{2021}\natexlab{}.
\newblock \bibinfo{title}{Engineering {AI} {Systems}: {A} {Research} {Agenda}}.
\newblock
\newblock
\urldef\tempurl%
\url{https://doi.org/10.4018/978-1-7998-5101-1.ch001}
\showDOI{\tempurl}
\newblock
\shownote{ISBN: 9781799851011 Pages: 1-19 Publisher: IGI Global.}


\bibitem[\protect\citeauthoryear{Buckland and Gey}{Buckland and Gey}{1994}]%
        {buckland_relationship_1994}
\bibfield{author}{\bibinfo{person}{Michael Buckland} {and}
  \bibinfo{person}{Fredric Gey}.} \bibinfo{year}{1994}\natexlab{}.
\newblock \showarticletitle{The relationship between recall and precision}.
\newblock \bibinfo{journal}{\emph{Journal of the American society for
  information science}} \bibinfo{volume}{45}, \bibinfo{number}{1}
  (\bibinfo{year}{1994}), \bibinfo{pages}{12--19}.
\newblock
\newblock
\shownote{Publisher: Wiley Online Library.}


\bibitem[\protect\citeauthoryear{Bürdek, Kehrer, Lochau, Reuling, Kelter, and
  Schürr}{Bürdek et~al\mbox{.}}{2016}]%
        {burdek_reasoning_2016}
\bibfield{author}{\bibinfo{person}{Johannes Bürdek}, \bibinfo{person}{Timo
  Kehrer}, \bibinfo{person}{Malte Lochau}, \bibinfo{person}{Dennis Reuling},
  \bibinfo{person}{Udo Kelter}, {and} \bibinfo{person}{Andy Schürr}.}
  \bibinfo{year}{2016}\natexlab{}.
\newblock \showarticletitle{Reasoning about product-line evolution using
  complex feature model differences}.
\newblock \bibinfo{journal}{\emph{Automated Software Engineering}}
  \bibinfo{volume}{23}, \bibinfo{number}{4} (\bibinfo{date}{Dec.}
  \bibinfo{year}{2016}), \bibinfo{pages}{687--733}.
\newblock
\showISSN{0928-8910, 1573-7535}
\urldef\tempurl%
\url{https://doi.org/10.1007/s10515-015-0185-3}
\showDOI{\tempurl}


\bibitem[\protect\citeauthoryear{Capilla, Bosch, Trinidad, Ruiz-Cortés, and
  Hinchey}{Capilla et~al\mbox{.}}{2014}]%
        {capilla_overview_2014}
\bibfield{author}{\bibinfo{person}{Rafael Capilla}, \bibinfo{person}{Jan
  Bosch}, \bibinfo{person}{Pablo Trinidad}, \bibinfo{person}{Antonio
  Ruiz-Cortés}, {and} \bibinfo{person}{Mike Hinchey}.}
  \bibinfo{year}{2014}\natexlab{}.
\newblock \showarticletitle{An overview of {Dynamic} {Software} {Product}
  {Line} architectures and techniques: {Observations} from research and
  industry}.
\newblock \bibinfo{journal}{\emph{Journal of Systems and Software}}
  \bibinfo{volume}{91} (\bibinfo{year}{2014}), \bibinfo{pages}{3--23}.
\newblock
\showISSN{0164-1212}
\urldef\tempurl%
\url{https://doi.org/10.1016/j.jss.2013.12.038}
\showDOI{\tempurl}


\bibitem[\protect\citeauthoryear{Carvalho, Pereira, and Cardoso}{Carvalho
  et~al\mbox{.}}{2019}]%
        {carvalho2019machine}
\bibfield{author}{\bibinfo{person}{Diogo~V Carvalho},
  \bibinfo{person}{Eduardo~M Pereira}, {and} \bibinfo{person}{Jaime~S
  Cardoso}.} \bibinfo{year}{2019}\natexlab{}.
\newblock \showarticletitle{Machine learning interpretability: A survey on
  methods and metrics}.
\newblock \bibinfo{journal}{\emph{Electronics}} \bibinfo{volume}{8},
  \bibinfo{number}{8} (\bibinfo{year}{2019}), \bibinfo{pages}{832}.
\newblock


\bibitem[\protect\citeauthoryear{Chen, Ali~Babar, and Nuseibeh}{Chen
  et~al\mbox{.}}{2013}]%
        {6365165}
\bibfield{author}{\bibinfo{person}{Lianping Chen}, \bibinfo{person}{Muhammad
  Ali~Babar}, {and} \bibinfo{person}{Bashar Nuseibeh}.}
  \bibinfo{year}{2013}\natexlab{}.
\newblock \showarticletitle{Characterizing Architecturally Significant
  Requirements}.
\newblock \bibinfo{journal}{\emph{IEEE Software}} \bibinfo{volume}{30},
  \bibinfo{number}{2} (\bibinfo{year}{2013}), \bibinfo{pages}{38--45}.
\newblock
\urldef\tempurl%
\url{https://doi.org/10.1109/MS.2012.174}
\showDOI{\tempurl}


\bibitem[\protect\citeauthoryear{Cook}{Cook}{1983}]%
        {cook1983overview}
\bibfield{author}{\bibinfo{person}{Stephen~A Cook}.}
  \bibinfo{year}{1983}\natexlab{}.
\newblock \showarticletitle{An overview of computational complexity}.
\newblock \bibinfo{journal}{\emph{Commun. ACM}} \bibinfo{volume}{26},
  \bibinfo{number}{6} (\bibinfo{year}{1983}), \bibinfo{pages}{400--408}.
\newblock


\bibitem[\protect\citeauthoryear{Cramer, Ford, and Hall}{Cramer
  et~al\mbox{.}}{1976}]%
        {cramer1976estimation}
\bibfield{author}{\bibinfo{person}{GM Cramer}, \bibinfo{person}{RA Ford}, {and}
  \bibinfo{person}{RL Hall}.} \bibinfo{year}{1976}\natexlab{}.
\newblock \showarticletitle{Estimation of toxic hazard—a decision tree
  approach}.
\newblock \bibinfo{journal}{\emph{Food and cosmetics toxicology}}
  \bibinfo{volume}{16}, \bibinfo{number}{3} (\bibinfo{year}{1976}),
  \bibinfo{pages}{255--276}.
\newblock


\bibitem[\protect\citeauthoryear{de~Souza~Nascimento, Ahmed, Oliveira, Palheta,
  Steinmacher, and Conte}{de~Souza~Nascimento et~al\mbox{.}}{2019}]%
        {de2019understanding}
\bibfield{author}{\bibinfo{person}{Elizamary de Souza~Nascimento},
  \bibinfo{person}{Iftekhar Ahmed}, \bibinfo{person}{Edson Oliveira},
  \bibinfo{person}{M{\'a}rcio~Piedade Palheta}, \bibinfo{person}{Igor
  Steinmacher}, {and} \bibinfo{person}{Tayana Conte}.}
  \bibinfo{year}{2019}\natexlab{}.
\newblock \showarticletitle{Understanding development process of machine
  learning systems: Challenges and solutions}. In
  \bibinfo{booktitle}{\emph{2019 ACM/IEEE International Symposium on Empirical
  Software Engineering and Measurement (ESEM)}}. IEEE, \bibinfo{pages}{1--6}.
\newblock


\bibitem[\protect\citeauthoryear{Di~Sipio, Di~Rocco, Di~Ruscio, and
  Nguyen}{Di~Sipio et~al\mbox{.}}{2021}]%
        {di_sipio_low-code_2021}
\bibfield{author}{\bibinfo{person}{Claudio Di~Sipio}, \bibinfo{person}{Juri
  Di~Rocco}, \bibinfo{person}{Davide Di~Ruscio}, {and} \bibinfo{person}{Dr.
  Phuong~Thanh Nguyen}.} \bibinfo{year}{2021}\natexlab{}.
\newblock \showarticletitle{A {Low}-{Code} {Tool} {Supporting} the
  {Development} of {Recommender} {Systems}}. In
  \bibinfo{booktitle}{\emph{Fifteenth {ACM} {Conference} on {Recommender}
  {Systems}}}. \bibinfo{publisher}{ACM}, \bibinfo{address}{Amsterdam
  Netherlands}, \bibinfo{pages}{741--744}.
\newblock
\showISBNx{978-1-4503-8458-2}
\urldef\tempurl%
\url{https://doi.org/10.1145/3460231.3478885}
\showDOI{\tempurl}


\bibitem[\protect\citeauthoryear{Domingos and Pazzani}{Domingos and
  Pazzani}{1997}]%
        {domingos_optimality_1997}
\bibfield{author}{\bibinfo{person}{Pedro Domingos} {and}
  \bibinfo{person}{Michael Pazzani}.} \bibinfo{year}{1997}\natexlab{}.
\newblock \showarticletitle{On the {Optimality} of the {Simple} {Bayesian}
  {Classifier} under {Zero}-{One} {Loss}}.
\newblock \bibinfo{journal}{\emph{Machine Learning}} \bibinfo{volume}{29},
  \bibinfo{number}{2} (\bibinfo{date}{Nov.} \bibinfo{year}{1997}),
  \bibinfo{pages}{103--130}.
\newblock
\showISSN{1573-0565}
\urldef\tempurl%
\url{https://doi.org/10.1023/A:1007413511361}
\showDOI{\tempurl}


\bibitem[\protect\citeauthoryear{Dwork, Hardt, Pitassi, Reingold, and
  Zemel}{Dwork et~al\mbox{.}}{2012}]%
        {dwork_fairness_2012}
\bibfield{author}{\bibinfo{person}{Cynthia Dwork}, \bibinfo{person}{Moritz
  Hardt}, \bibinfo{person}{Toniann Pitassi}, \bibinfo{person}{Omer Reingold},
  {and} \bibinfo{person}{Richard Zemel}.} \bibinfo{year}{2012}\natexlab{}.
\newblock \showarticletitle{Fairness through awareness}.
\newblock \bibinfo{journal}{\emph{Proceedings of the 3rd Innovations in
  Theoretical Computer Science Conference}} (\bibinfo{year}{2012}),
  \bibinfo{pages}{214--226}.
\newblock
\showISBNx{978-1-4503-1115-1}
\urldef\tempurl%
\url{https://doi.org/10.1145/2090236.2090255}
\showDOI{\tempurl}


\bibitem[\protect\citeauthoryear{Feldman, Friedler, Moeller, Scheidegger, and
  Venkatasubramanian}{Feldman et~al\mbox{.}}{2015}]%
        {feldman_certifying_2015}
\bibfield{author}{\bibinfo{person}{Michael Feldman},
  \bibinfo{person}{Sorelle~A. Friedler}, \bibinfo{person}{John Moeller},
  \bibinfo{person}{Carlos Scheidegger}, {and} \bibinfo{person}{Suresh
  Venkatasubramanian}.} \bibinfo{year}{2015}\natexlab{}.
\newblock \showarticletitle{Certifying and {Removing} {Disparate} {Impact}}. In
  \bibinfo{booktitle}{\emph{Proceedings of the 21th {ACM} {SIGKDD}
  {International} {Conference} on {Knowledge} {Discovery} and {Data}
  {Mining}}}. \bibinfo{publisher}{ACM}, \bibinfo{address}{Sydney NSW
  Australia}, \bibinfo{pages}{259--268}.
\newblock
\showISBNx{978-1-4503-3664-2}
\urldef\tempurl%
\url{https://doi.org/10.1145/2783258.2783311}
\showDOI{\tempurl}


\bibitem[\protect\citeauthoryear{Fung, Wang, Chen, and Yu}{Fung
  et~al\mbox{.}}{2010}]%
        {10.1145/1749603.1749605}
\bibfield{author}{\bibinfo{person}{Benjamin C.~M. Fung}, \bibinfo{person}{Ke
  Wang}, \bibinfo{person}{Rui Chen}, {and} \bibinfo{person}{Philip~S. Yu}.}
  \bibinfo{year}{2010}\natexlab{}.
\newblock \showarticletitle{Privacy-Preserving Data Publishing: A Survey of
  Recent Developments}.
\newblock \bibinfo{journal}{\emph{ACM Comput. Surv.}} \bibinfo{volume}{42},
  \bibinfo{number}{4}, Article \bibinfo{articleno}{14} (\bibinfo{date}{June}
  \bibinfo{year}{2010}), \bibinfo{numpages}{53}~pages.
\newblock
\showISSN{0360-0300}
\urldef\tempurl%
\url{https://doi.org/10.1145/1749603.1749605}
\showDOI{\tempurl}


\bibitem[\protect\citeauthoryear{Giray}{Giray}{2021}]%
        {giray2021software}
\bibfield{author}{\bibinfo{person}{G{\"o}rkem Giray}.}
  \bibinfo{year}{2021}\natexlab{}.
\newblock \showarticletitle{A software engineering perspective on engineering
  machine learning systems: State of the art and challenges}.
\newblock \bibinfo{journal}{\emph{Journal of Systems and Software}}
  (\bibinfo{year}{2021}), \bibinfo{pages}{111031}.
\newblock


\bibitem[\protect\citeauthoryear{{Goncalves Jr.} and {Barros}}{{Goncalves Jr.}
  and {Barros}}{2011}]%
        {6086455}
\bibfield{author}{\bibinfo{person}{P.~M. {Goncalves Jr.}} {and}
  \bibinfo{person}{R.~S.~M. {Barros}}.} \bibinfo{year}{2011}\natexlab{}.
\newblock \showarticletitle{Automating Data Preprocessing with DMPML and
  KDDML}. In \bibinfo{booktitle}{\emph{2011 10th IEEE/ACIS International
  Conference on Computer and Information Science}}. \bibinfo{pages}{97--103}.
\newblock
\urldef\tempurl%
\url{https://doi.org/10.1109/ICIS.2011.23}
\showDOI{\tempurl}


\bibitem[\protect\citeauthoryear{Guo, Wang, Bell, Bi, and Greer}{Guo
  et~al\mbox{.}}{2003}]%
        {guo2003knn}
\bibfield{author}{\bibinfo{person}{Gongde Guo}, \bibinfo{person}{Hui Wang},
  \bibinfo{person}{David Bell}, \bibinfo{person}{Yaxin Bi}, {and}
  \bibinfo{person}{Kieran Greer}.} \bibinfo{year}{2003}\natexlab{}.
\newblock \showarticletitle{KNN model-based approach in classification}. In
  \bibinfo{booktitle}{\emph{OTM Confederated International Conferences" On the
  Move to Meaningful Internet Systems"}}. Springer, \bibinfo{pages}{986--996}.
\newblock


\bibitem[\protect\citeauthoryear{Hagan, Demuth, and Beale}{Hagan
  et~al\mbox{.}}{1997}]%
        {hagan1997neural}
\bibfield{author}{\bibinfo{person}{Martin~T Hagan}, \bibinfo{person}{Howard~B
  Demuth}, {and} \bibinfo{person}{Mark Beale}.}
  \bibinfo{year}{1997}\natexlab{}.
\newblock \bibinfo{booktitle}{\emph{Neural network design}}.
\newblock \bibinfo{publisher}{PWS Publishing Co.}
\newblock


\bibitem[\protect\citeauthoryear{Hamada, Ishikawa, Masuda, Myojin, Nishi,
  Ogawa, Toku, Tokumoto, Tsuchiya, Ujita, et~al\mbox{.}}{Hamada
  et~al\mbox{.}}{2020}]%
        {hamada2020guidelines}
\bibfield{author}{\bibinfo{person}{Koichi Hamada}, \bibinfo{person}{Fuyuki
  Ishikawa}, \bibinfo{person}{Satoshi Masuda}, \bibinfo{person}{Tomoyuki
  Myojin}, \bibinfo{person}{Yasuharu Nishi}, \bibinfo{person}{Hideto Ogawa},
  \bibinfo{person}{Takahiro Toku}, \bibinfo{person}{Susumu Tokumoto},
  \bibinfo{person}{Kazunori Tsuchiya}, \bibinfo{person}{Yasuhiro Ujita},
  {et~al\mbox{.}}} \bibinfo{year}{2020}\natexlab{}.
\newblock \showarticletitle{Guidelines for Quality Assurance of Machine
  Learning-based Artificial Intelligence.}. In
  \bibinfo{booktitle}{\emph{SEKE}}. \bibinfo{pages}{335--341}.
\newblock


\bibitem[\protect\citeauthoryear{Hapke and Nelson}{Hapke and Nelson}{2020}]%
        {hapke_building_2020}
\bibfield{author}{\bibinfo{person}{Hannes Hapke} {and}
  \bibinfo{person}{Catherine Nelson}.} \bibinfo{year}{2020}\natexlab{}.
\newblock \bibinfo{booktitle}{\emph{Building {Machine} {Learning}
  {Pipelines}}}.
\newblock \bibinfo{publisher}{"O'Reilly Media, Inc."}.
\newblock
\showISBNx{978-1-4920-5316-3}
\newblock
\shownote{Google-Books-ID: H6\_wDwAAQBAJ.}


\bibitem[\protect\citeauthoryear{Hardt, Price, Price, and Srebro}{Hardt
  et~al\mbox{.}}{2016}]%
        {hardt_equality_2016}
\bibfield{author}{\bibinfo{person}{Moritz Hardt}, \bibinfo{person}{Eric Price},
  \bibinfo{person}{Eric Price}, {and} \bibinfo{person}{Nati Srebro}.}
  \bibinfo{year}{2016}\natexlab{}.
\newblock \showarticletitle{Equality of {Opportunity} in {Supervised}
  {Learning}}. In \bibinfo{booktitle}{\emph{Advances in {Neural} {Information}
  {Processing} {Systems}}}, Vol.~\bibinfo{volume}{29}.
  \bibinfo{publisher}{Curran Associates, Inc.}
\newblock
\urldef\tempurl%
\url{https://proceedings.neurips.cc/paper/2016/hash/9d2682367c3935defcb1f9e247a97c0d-Abstract.html}
\showURL{%
\tempurl}


\bibitem[\protect\citeauthoryear{He, Zhao, and Chu}{He et~al\mbox{.}}{2021}]%
        {HE2021106622}
\bibfield{author}{\bibinfo{person}{Xin He}, \bibinfo{person}{Kaiyong Zhao},
  {and} \bibinfo{person}{Xiaowen Chu}.} \bibinfo{year}{2021}\natexlab{}.
\newblock \showarticletitle{AutoML: A survey of the state-of-the-art}.
\newblock \bibinfo{journal}{\emph{Knowledge-Based Systems}}
  \bibinfo{volume}{212} (\bibinfo{year}{2021}), \bibinfo{pages}{106622}.
\newblock
\showISSN{0950-7051}
\urldef\tempurl%
\url{https://doi.org/10.1016/j.knosys.2020.106622}
\showDOI{\tempurl}


\bibitem[\protect\citeauthoryear{Ishikawa}{Ishikawa}{2018}]%
        {ishikawa2018concepts}
\bibfield{author}{\bibinfo{person}{Fuyuki Ishikawa}.}
  \bibinfo{year}{2018}\natexlab{}.
\newblock \showarticletitle{Concepts in quality assessment for machine
  learning-from test data to arguments}. In
  \bibinfo{booktitle}{\emph{International Conference on Conceptual Modeling}}.
  Springer, \bibinfo{pages}{536--544}.
\newblock


\bibitem[\protect\citeauthoryear{ISO}{ISO}{2011}]%
        {iso_isoiec_nodate}
\bibfield{author}{\bibinfo{person}{ISO}.} \bibinfo{year}{2011}\natexlab{}.
\newblock \bibinfo{booktitle}{\emph{{ISO}/{IEC} 25010:2011}}.
\newblock \bibinfo{type}{{T}echnical {R}eport}.
\newblock
\urldef\tempurl%
\url{https://www.iso.org/cms/render/live/en/sites/isoorg/contents/data/standard/03/57/35733.html}
\showURL{%
\tempurl}


\bibitem[\protect\citeauthoryear{Kamiran and Calders}{Kamiran and
  Calders}{2012}]%
        {kamiran_data_2012}
\bibfield{author}{\bibinfo{person}{Faisal Kamiran} {and} \bibinfo{person}{Toon
  Calders}.} \bibinfo{year}{2012}\natexlab{}.
\newblock \showarticletitle{Data preprocessing techniques for classification
  without discrimination}.
\newblock \bibinfo{journal}{\emph{Knowledge and Information Systems}}
  \bibinfo{volume}{33}, \bibinfo{number}{1} (\bibinfo{date}{Oct.}
  \bibinfo{year}{2012}), \bibinfo{pages}{1--33}.
\newblock
\showISSN{0219-1377, 0219-3116}
\urldef\tempurl%
\url{https://doi.org/10.1007/s10115-011-0463-8}
\showDOI{\tempurl}


\bibitem[\protect\citeauthoryear{Kang, Cohen, Hess, Novak, and Peterson}{Kang
  et~al\mbox{.}}{1990}]%
        {kang1990feature}
\bibfield{author}{\bibinfo{person}{Kyo~C Kang}, \bibinfo{person}{Sholom~G
  Cohen}, \bibinfo{person}{James~A Hess}, \bibinfo{person}{William~E Novak},
  {and} \bibinfo{person}{A~Spencer Peterson}.} \bibinfo{year}{1990}\natexlab{}.
\newblock \bibinfo{booktitle}{\emph{Feature-oriented domain analysis (FODA)
  feasibility study}}.
\newblock \bibinfo{type}{{T}echnical {R}eport}.
  \bibinfo{institution}{Carnegie-Mellon Univ Pittsburgh Pa Software Engineering
  Inst}.
\newblock


\bibitem[\protect\citeauthoryear{Kearns}{Kearns}{1990}]%
        {kearns1990computational}
\bibfield{author}{\bibinfo{person}{Michael~J Kearns}.}
  \bibinfo{year}{1990}\natexlab{}.
\newblock \bibinfo{booktitle}{\emph{The computational complexity of machine
  learning}}.
\newblock \bibinfo{publisher}{MIT press}.
\newblock


\bibitem[\protect\citeauthoryear{Kumeno}{Kumeno}{2019}]%
        {kumeno2019sofware}
\bibfield{author}{\bibinfo{person}{Fumihiro Kumeno}.}
  \bibinfo{year}{2019}\natexlab{}.
\newblock \showarticletitle{Sofware engneering challenges for machine learning
  applications: A literature review}.
\newblock \bibinfo{journal}{\emph{Intelligent Decision Technologies}}
  \bibinfo{volume}{13}, \bibinfo{number}{4} (\bibinfo{year}{2019}),
  \bibinfo{pages}{463--476}.
\newblock


\bibitem[\protect\citeauthoryear{Kusner, Loftus, Russell, and Silva}{Kusner
  et~al\mbox{.}}{2017}]%
        {kusner_counterfactual_2017}
\bibfield{author}{\bibinfo{person}{Matt~J Kusner}, \bibinfo{person}{Joshua
  Loftus}, \bibinfo{person}{Chris Russell}, {and} \bibinfo{person}{Ricardo
  Silva}.} \bibinfo{year}{2017}\natexlab{}.
\newblock \showarticletitle{Counterfactual Fairness}. In
  \bibinfo{booktitle}{\emph{Advances in Neural Information Processing Systems}}
  (2017), Vol.~\bibinfo{volume}{30}. \bibinfo{publisher}{Curran Associates,
  Inc.}
\newblock
\urldef\tempurl%
\url{https://proceedings.neurips.cc/paper/2017/hash/a486cd07e4ac3d270571622f4f316ec5-Abstract.html}
\showURL{%
\tempurl}


\bibitem[\protect\citeauthoryear{Linardatos, Papastefanopoulos, and
  Kotsiantis}{Linardatos et~al\mbox{.}}{2021}]%
        {linardatos2021explainable}
\bibfield{author}{\bibinfo{person}{Pantelis Linardatos},
  \bibinfo{person}{Vasilis Papastefanopoulos}, {and} \bibinfo{person}{Sotiris
  Kotsiantis}.} \bibinfo{year}{2021}\natexlab{}.
\newblock \showarticletitle{Explainable AI: A Review of Machine Learning
  Interpretability Methods}.
\newblock \bibinfo{journal}{\emph{Entropy}} \bibinfo{volume}{23},
  \bibinfo{number}{1} (\bibinfo{year}{2021}), \bibinfo{pages}{18}.
\newblock


\bibitem[\protect\citeauthoryear{Mart{\'\i}nez-Plumed, Contreras-Ochando,
  Ferri, Orallo, Kull, Lachiche, Quintana, and Flach}{Mart{\'\i}nez-Plumed
  et~al\mbox{.}}{2019}]%
        {martinez2019crisp}
\bibfield{author}{\bibinfo{person}{Fernando Mart{\'\i}nez-Plumed},
  \bibinfo{person}{Lidia Contreras-Ochando}, \bibinfo{person}{Cesar Ferri},
  \bibinfo{person}{Jose~Hernandez Orallo}, \bibinfo{person}{Meelis Kull},
  \bibinfo{person}{Nicolas Lachiche}, \bibinfo{person}{Mar{\'e}a
  Jos{\'e}~Ram{\'\i}rez Quintana}, {and} \bibinfo{person}{Peter~A Flach}.}
  \bibinfo{year}{2019}\natexlab{}.
\newblock \showarticletitle{CRISP-DM twenty years later: From data mining
  processes to data science trajectories}.
\newblock \bibinfo{journal}{\emph{IEEE Transactions on Knowledge and Data
  Engineering}} (\bibinfo{year}{2019}).
\newblock


\bibitem[\protect\citeauthoryear{Martínez-Fernández, Bogner, Franch, Oriol,
  Siebert, Trendowicz, Vollmer, and Wagner}{Martínez-Fernández
  et~al\mbox{.}}{2022}]%
        {martinez-fernandez_software_2022}
\bibfield{author}{\bibinfo{person}{Silverio Martínez-Fernández},
  \bibinfo{person}{Justus Bogner}, \bibinfo{person}{Xavier Franch},
  \bibinfo{person}{Marc Oriol}, \bibinfo{person}{Julien Siebert},
  \bibinfo{person}{Adam Trendowicz}, \bibinfo{person}{Anna~Maria Vollmer},
  {and} \bibinfo{person}{Stefan Wagner}.} \bibinfo{year}{2022}\natexlab{}.
\newblock \showarticletitle{Software {Engineering} for {AI}-{Based} {Systems}:
  {A} {Survey}}.
\newblock \bibinfo{journal}{\emph{ACM Transactions on Software Engineering and
  Methodology}} \bibinfo{volume}{31}, \bibinfo{number}{2}
  (\bibinfo{date}{April} \bibinfo{year}{2022}), \bibinfo{pages}{1--59}.
\newblock
\showISSN{1049-331X, 1557-7392}
\urldef\tempurl%
\url{https://doi.org/10.1145/3487043}
\showDOI{\tempurl}
\newblock
\shownote{arXiv:2105.01984 [cs].}


\bibitem[\protect\citeauthoryear{Mehrabi, Morstatter, Saxena, Lerman, and
  Galstyan}{Mehrabi et~al\mbox{.}}{2021}]%
        {mehrabi_survey_2021}
\bibfield{author}{\bibinfo{person}{Ninareh Mehrabi}, \bibinfo{person}{Fred
  Morstatter}, \bibinfo{person}{Nripsuta Saxena}, \bibinfo{person}{Kristina
  Lerman}, {and} \bibinfo{person}{Aram Galstyan}.}
  \bibinfo{year}{2021}\natexlab{}.
\newblock \showarticletitle{A {Survey} on {Bias} and {Fairness} in {Machine}
  {Learning}}.
\newblock \bibinfo{journal}{\emph{Comput. Surveys}} \bibinfo{volume}{54},
  \bibinfo{number}{6} (\bibinfo{date}{July} \bibinfo{year}{2021}),
  \bibinfo{pages}{1--35}.
\newblock
\showISSN{0360-0300, 1557-7341}
\urldef\tempurl%
\url{https://doi.org/10.1145/3457607}
\showDOI{\tempurl}


\bibitem[\protect\citeauthoryear{Menard}{Menard}{2002}]%
        {menard2002applied}
\bibfield{author}{\bibinfo{person}{Scott Menard}.}
  \bibinfo{year}{2002}\natexlab{}.
\newblock \bibinfo{booktitle}{\emph{Applied logistic regression analysis}}.
  Vol.~\bibinfo{volume}{106}.
\newblock \bibinfo{publisher}{Sage}.
\newblock


\bibitem[\protect\citeauthoryear{Muccini and Vaidhyanathan}{Muccini and
  Vaidhyanathan}{2021}]%
        {muccini_software_2021}
\bibfield{author}{\bibinfo{person}{Henry Muccini} {and}
  \bibinfo{person}{Karthik Vaidhyanathan}.} \bibinfo{year}{2021}\natexlab{}.
\newblock \showarticletitle{Software {Architecture} for {ML}-based {Systems}:
  {What} {Exists} and {What} {Lies} {Ahead}}. In \bibinfo{booktitle}{\emph{2021
  {IEEE}/{ACM} 1st {Workshop} on {AI} {Engineering} - {Software} {Engineering}
  for {AI} ({WAIN})}}. \bibinfo{pages}{121--128}.
\newblock
\urldef\tempurl%
\url{https://doi.org/10.1109/WAIN52551.2021.00026}
\showDOI{\tempurl}


\bibitem[\protect\citeauthoryear{Nations}{Nations}{[n.d.]}]%
        {united_nations_17_nodate}
\bibfield{author}{\bibinfo{person}{United Nations}.}
  \bibinfo{year}{[n.d.]}\natexlab{}.
\newblock \bibinfo{title}{{THE} 17 {GOALS} {\textbar} {Sustainable}
  {Development}}.
\newblock
\newblock
\urldef\tempurl%
\url{https://sdgs.un.org/goals}
\showURL{%
\tempurl}


\bibitem[\protect\citeauthoryear{Putzel and Lee}{Putzel and Lee}{2022}]%
        {putzel_blackbox_2022}
\bibfield{author}{\bibinfo{person}{Preston Putzel} {and} \bibinfo{person}{Scott
  Lee}.} \bibinfo{year}{2022}\natexlab{}.
\newblock \showarticletitle{Blackbox {Post}-{Processing} for {Multiclass}
  {Fairness}}.
\newblock \bibinfo{journal}{\emph{arXiv:2201.04461 [cs]}} (\bibinfo{date}{Jan.}
  \bibinfo{year}{2022}).
\newblock
\urldef\tempurl%
\url{http://arxiv.org/abs/2201.04461}
\showURL{%
\tempurl}
\newblock
\shownote{arXiv: 2201.04461.}


\bibitem[\protect\citeauthoryear{Rosenfield and Fitzpatrick-Lins}{Rosenfield
  and Fitzpatrick-Lins}{1986}]%
        {rosenfield_coefficient_1986}
\bibfield{author}{\bibinfo{person}{G.H. Rosenfield} {and} \bibinfo{person}{K.
  Fitzpatrick-Lins}.} \bibinfo{year}{1986}\natexlab{}.
\newblock \showarticletitle{A coefficient of agreement as a measure of thematic
  classification accuracy.}
\newblock \bibinfo{journal}{\emph{Photogrammetric Engineering and Remote
  Sensing}} \bibinfo{volume}{52}, \bibinfo{number}{2} (\bibinfo{year}{1986}),
  \bibinfo{pages}{223--227}.
\newblock
\urldef\tempurl%
\url{http://pubs.er.usgs.gov/publication/70014667}
\showURL{%
\tempurl}


\bibitem[\protect\citeauthoryear{Ruehle, Sim, Yekhanin, Chandran, Chase, Jones,
  Laine, Kopf, Teevan, Kleewein, and Rajmohan}{Ruehle et~al\mbox{.}}{2021}]%
        {ruehle_privacy_2021}
\bibfield{author}{\bibinfo{person}{Victor Ruehle}, \bibinfo{person}{Robert
  Sim}, \bibinfo{person}{Sergey Yekhanin}, \bibinfo{person}{Nishanth Chandran},
  \bibinfo{person}{Melissa Chase}, \bibinfo{person}{Daniel Jones},
  \bibinfo{person}{Kim Laine}, \bibinfo{person}{Boris Kopf},
  \bibinfo{person}{James Teevan}, \bibinfo{person}{Jim Kleewein}, {and}
  \bibinfo{person}{Saravan Rajmohan}.} \bibinfo{year}{2021}\natexlab{}.
\newblock \bibinfo{title}{Privacy {Preserving} {Machine} {Learning}:
  {Maintaining} confidentiality and preserving trust}.
\newblock
\newblock
\urldef\tempurl%
\url{https://www.microsoft.com/en-us/research/blog/privacy-preserving-machine-learning-maintaining-confidentiality-and-preserving-trust/}
\showURL{%
\tempurl}


\bibitem[\protect\citeauthoryear{Rönkkö, Heikkinen, Kotovirta, and
  Chandrasekar}{Rönkkö et~al\mbox{.}}{2015}]%
        {RONKKO201513}
\bibfield{author}{\bibinfo{person}{Mauno Rönkkö}, \bibinfo{person}{Jani
  Heikkinen}, \bibinfo{person}{Ville Kotovirta}, {and}
  \bibinfo{person}{Venkatachalam Chandrasekar}.}
  \bibinfo{year}{2015}\natexlab{}.
\newblock \showarticletitle{Automated preprocessing of environmental data}.
\newblock \bibinfo{journal}{\emph{Future Generation Computer Systems}}
  \bibinfo{volume}{45} (\bibinfo{year}{2015}), \bibinfo{pages}{13--24}.
\newblock
\showISSN{0167-739X}
\urldef\tempurl%
\url{https://doi.org/10.1016/j.future.2014.10.011}
\showDOI{\tempurl}


\bibitem[\protect\citeauthoryear{Siebert, Joeckel, Heidrich, Trendowicz,
  Nakamichi, Ohashi, Namba, Yamamoto, and Aoyama}{Siebert
  et~al\mbox{.}}{2021}]%
        {siebert2021construction}
\bibfield{author}{\bibinfo{person}{Julien Siebert}, \bibinfo{person}{Lisa
  Joeckel}, \bibinfo{person}{Jens Heidrich}, \bibinfo{person}{Adam Trendowicz},
  \bibinfo{person}{Koji Nakamichi}, \bibinfo{person}{Kyoko Ohashi},
  \bibinfo{person}{Isao Namba}, \bibinfo{person}{Rieko Yamamoto}, {and}
  \bibinfo{person}{Mikio Aoyama}.} \bibinfo{year}{2021}\natexlab{}.
\newblock \showarticletitle{Construction of a quality model for machine
  learning systems}.
\newblock \bibinfo{journal}{\emph{Software Quality Journal}}
  (\bibinfo{year}{2021}), \bibinfo{pages}{1--29}.
\newblock


\bibitem[\protect\citeauthoryear{Steinberg, Budinsky, Paternostro, and
  Merks}{Steinberg et~al\mbox{.}}{2008}]%
        {steinberg_emf_2008}
\bibfield{author}{\bibinfo{person}{Dave Steinberg}, \bibinfo{person}{Frank
  Budinsky}, \bibinfo{person}{Marcelo Paternostro}, {and} \bibinfo{person}{Ed
  Merks}.} \bibinfo{year}{2008}\natexlab{}.
\newblock \bibinfo{booktitle}{\emph{{EMF}: {Eclipse} {Modeling} {Framework},
  2nd {Edition}} (\bibinfo{edition}{2nd} ed.)}.
\newblock \bibinfo{publisher}{Addison-Wesley Professional.}
\newblock
\showISBNx{978-0-321-33188-5}
\urldef\tempurl%
\url{https://www.informit.com/store/emf-eclipse-modeling-framework-9780321331885}
\showURL{%
\tempurl}


\bibitem[\protect\citeauthoryear{Studer, Bui, Drescher, Hanuschkin, Winkler,
  Peters, and M{\"u}ller}{Studer et~al\mbox{.}}{2021}]%
        {studer2021towards}
\bibfield{author}{\bibinfo{person}{Stefan Studer}, \bibinfo{person}{Thanh~Binh
  Bui}, \bibinfo{person}{Christian Drescher}, \bibinfo{person}{Alexander
  Hanuschkin}, \bibinfo{person}{Ludwig Winkler}, \bibinfo{person}{Steven
  Peters}, {and} \bibinfo{person}{Klaus-Robert M{\"u}ller}.}
  \bibinfo{year}{2021}\natexlab{}.
\newblock \showarticletitle{Towards CRISP-ML (Q): a machine learning process
  model with quality assurance methodology}.
\newblock \bibinfo{journal}{\emph{Machine Learning and Knowledge Extraction}}
  \bibinfo{volume}{3}, \bibinfo{number}{2} (\bibinfo{year}{2021}),
  \bibinfo{pages}{392--413}.
\newblock


\bibitem[\protect\citeauthoryear{Taha and Hanbury}{Taha and Hanbury}{2015}]%
        {taha_metrics_2015}
\bibfield{author}{\bibinfo{person}{Abdel~Aziz Taha} {and}
  \bibinfo{person}{Allan Hanbury}.} \bibinfo{year}{2015}\natexlab{}.
\newblock \showarticletitle{Metrics for evaluating {3D} medical image
  segmentation: analysis, selection, and tool}.
\newblock \bibinfo{journal}{\emph{BMC Medical Imaging}} \bibinfo{volume}{15},
  \bibinfo{number}{1} (\bibinfo{date}{Aug.} \bibinfo{year}{2015}),
  \bibinfo{pages}{29}.
\newblock
\showISSN{1471-2342}
\urldef\tempurl%
\url{https://doi.org/10.1186/s12880-015-0068-x}
\showDOI{\tempurl}


\bibitem[\protect\citeauthoryear{Thum, Kastner, Erdweg, and Siegmund}{Thum
  et~al\mbox{.}}{2011}]%
        {thum2011abstract}
\bibfield{author}{\bibinfo{person}{Thomas Thum}, \bibinfo{person}{Christian
  Kastner}, \bibinfo{person}{Sebastian Erdweg}, {and} \bibinfo{person}{Norbert
  Siegmund}.} \bibinfo{year}{2011}\natexlab{}.
\newblock \showarticletitle{Abstract features in feature modeling}. In
  \bibinfo{booktitle}{\emph{2011 15th International Software Product Line
  Conference}}. IEEE, \bibinfo{pages}{191--200}.
\newblock


\bibitem[\protect\citeauthoryear{Villamizar, Escovedo, and
  Kalinowski}{Villamizar et~al\mbox{.}}{2021}]%
        {villamizarrequirements}
\bibfield{author}{\bibinfo{person}{Hugo Villamizar}, \bibinfo{person}{Tatiana
  Escovedo}, {and} \bibinfo{person}{Marcos Kalinowski}.}
  \bibinfo{year}{2021}\natexlab{}.
\newblock \showarticletitle{Requirements Engineering for Machine Learning: A
  Systematic Mapping Study}. In \bibinfo{booktitle}{\emph{SEAA}}.
  \bibinfo{pages}{29--36}.
\newblock


\bibitem[\protect\citeauthoryear{Viyović, Maksimović, and Perisić}{Viyović
  et~al\mbox{.}}{2014}]%
        {viyovic_sirius_2014}
\bibfield{author}{\bibinfo{person}{Vladimir Viyović}, \bibinfo{person}{Mirjam
  Maksimović}, {and} \bibinfo{person}{Branko Perisić}.}
  \bibinfo{year}{2014}\natexlab{}.
\newblock \showarticletitle{Sirius: {A} rapid development of {DSM} graphical
  editor}. In \bibinfo{booktitle}{\emph{{IEEE} 18th {International}
  {Conference} on {Intelligent} {Engineering} {Systems} {INES} 2014}}.
  \bibinfo{pages}{233--238}.
\newblock
\urldef\tempurl%
\url{https://doi.org/10.1109/INES.2014.6909375}
\showDOI{\tempurl}
\newblock
\shownote{ISSN: 1543-9259.}


\bibitem[\protect\citeauthoryear{Zhang, Harman, Ma, and Liu}{Zhang
  et~al\mbox{.}}{2020}]%
        {zhang2020machine}
\bibfield{author}{\bibinfo{person}{Jie~M Zhang}, \bibinfo{person}{Mark Harman},
  \bibinfo{person}{Lei Ma}, {and} \bibinfo{person}{Yang Liu}.}
  \bibinfo{year}{2020}\natexlab{}.
\newblock \showarticletitle{Machine learning testing: Survey, landscapes and
  horizons}.
\newblock \bibinfo{journal}{\emph{IEEE Transactions on Software Engineering}}
  (\bibinfo{year}{2020}).
\newblock


\end{thebibliography}

\end{document}